# *CrowdEgress: A Multi-Agent Simulation Platform for Pedestrian Crowd*

**Peng Wang Xiaoda Wang Peter Luh**

**Neal Olderman Christian Wilkie Timo Korhonen**

*This manual introduces a simulation tool to study complex crowd behavior in social context. The agent-based model is extended based on the well-known social force model, and it mainly describes how agents interact with each other, and also with surrounding facilities such as walls, doors and exits. The simulation platform is compatible to FDS+Evac, and the input data in FDS+Evac could be imported into our simulation platform to create single-floor compartment geometry, and a flow solver is used to generate the road map towards exits. Most importantly, we plan to integrate advanced social and psychological theory into our simulation platform, especially investigating human behavior in emergency evacuation, such as pre-evacuation behavior, exit-selection activities, social group and herding effect and so forth*[1]*.*

## *1. Introduction*

*The agent-based model (ABM) is a computational research method to study social systems. This model-driven approach partly origins from statistical physics, investigating how individuals in free space or in lattice interact with each other and whether there is any converged pattern emerged at the macroscopic level. Since a society composed of many people is a typical system of many-particle, it is possible to apply the principles of statistical physics to study social behavior of many individuals. Recently, there has been a growing interest in using agent-based model and simulation to understand social phenomena such as economic market or political opinions [Peralta, Kertesz, Iniguez, 2022; Quang et. al., 2018].*

*The agent-based model refers to a system of many-particle that exhibits emergent characteristics when autonomous agents interact with each other. Basically, the ABM consists of agent, system space, and external environment. The agents are autonomous and decides their behavior by interacting with the neighbors or the external environment with the rules of behavior. In our simulation platform, for example, the system space is a 2D planar space, and the environment is given as a structural layout that consists of obstructions (e.g., walls) and passageways (e.g., doors or exits), and other environmental stimuli may be imported such as gas temperature or smoke density in the future. Agents are interacting with each other and moving within this structural layout.*

*Very importantly, the individual agent is formulated by merging knowledge in broad range from statistical physics to sociology and dynamical systems, and the macroscopic phenomena emerging from individual interactions are studied by numerical methods. The agent-based model and simulation mainly aims at study of collective behavior of human pedestrian crowd, and it could also be applied for simulation of other socialized animals like bird flock, fish school, or sheep herd.*

*The simulation is mainly implemented by a component as packed in a class called simulation class (simulation.py), and it computes interaction of agents with surrounding entities including walls, doors and exits. The agent model is described in agent.py, whereas walls, doors and exits are coded in obst.py. The agent-based model is an extension of the well-known social force model (Helbing and Monlar 1995; Helbing, Farkas, Vicsek 2000; Lakoba, Kaup and Finkelstein 2005). This model has been applied in many existing pedestrian simulators such as PTV Viswalk,*

[1] The research program funded by NSF Grant # CMMI-1000495 (NSF Program Name: Building Emergency Evacuation - Innovative Modeling and Optimization).

*MassMotion, FDS+Evac and so forth [Santos and Aguirre 2005; Ronchi, R. Lovreglio, M. Kinsey, 2020]. The model aims at investigating prototypes of pedestrian behavior in crowd evacuation. The core algorithm is still being developed and improved. This is an inter-discipline study topic, which refers to Newton particles, statistical physics, dynamic systems as well as social and behavioral science. Your contributions or comments are much welcome. The program source code and numerical test cases are mainly uploaded online at https://github.com/godisreal/crowdEgress and https://sourceforge.net/p/crowdegress/code*

*The program also consists of several functional components such as User Interface and Data/Visualization Tool.*

***User Interface:*** *The user interface is written in tkinter in ui.py. Please run ui.py to enable a graphic user interface (GUI) where one selects the input files, initialize compartment geometry, and configure or start a simulation. An alternative method is using main.py to directly start a simulation without GUI. Currently there is a simple version of GUI and it needs to be improved in several aspects.*

***Data Tool****: This component is packed in data_func.py, and it reads in data from input files, and write data to output files. The input data is written by users in either csv files or fds input files [McGrattan et. al., 2021]. Agents must be specified in csv file while walls, doors or exits can be described either in csv file or read from standard fds input file. The simulation output is written into a binary file, which is compatible to the fds output data (fds prt5 data format).*

*In the future we plan to visualize such data by smokeview [Forney, 2022], which is the standard tool to visualize fds output data.*

***Visualization Tool:*** *The visualization component is packed in draw_func.py and currently pygame (SDL for python) is used to develop this component. Users can select to visualize the simulation as it runs, or visualize the output data after the simulation is complete. If anyone is interested, please feel free to extend the module or try other graphic libraries to write a visualization component.*

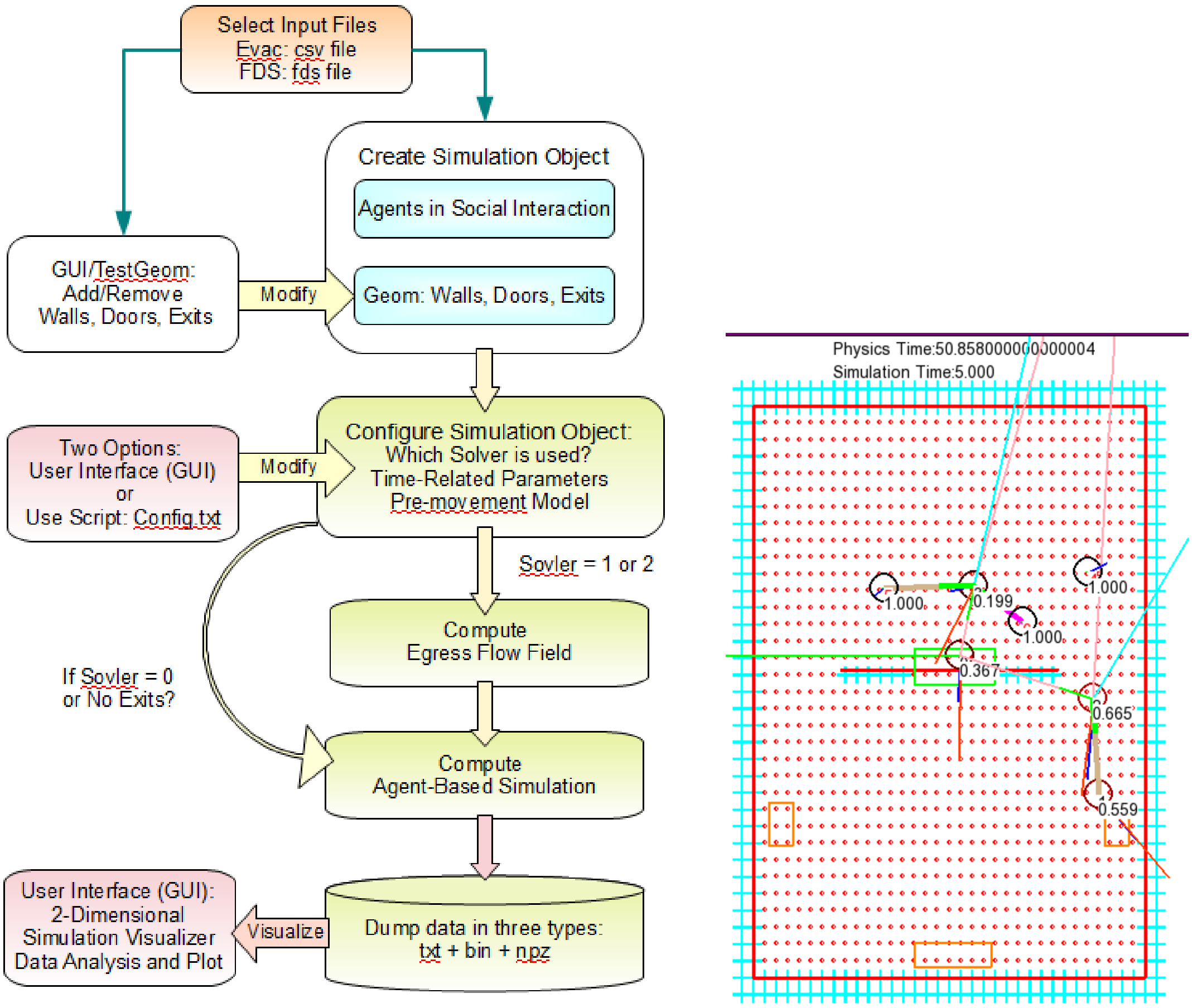

***Figure 1. Program Flow Chart.***

*As above we present the program flow chart in Figure 1. The input data must include a csv file to describe agents in the social context. The compartment data such as walls, doors and exits are either included in the csv file or created by a FDS input file. The output data files include a binary data file, a text file and a npz data file (a special form in Numpy to store data array or matrix-like data). The solver for agent-based simulation is listed as below.*

***Table 1. Solver of Agent-Based Model***

| | |
|---|---|
| *Solver = 2* | *Compute egress flow field for all possible exits (See Figure 8) and agents select one exit based on exit-selection probability. Agents move to a selected exit when simulation time exceeds his/her pre-movement time .* |
| *Solver = 1* | *Compute egress flow field for the nearest-exit strategy (See Figure 7), and agents follow this flow field to use the nearest-exit to their locations. Agents move to the nearest exit when simulation time exceeds his/her pre-movement time .* |
| *Solver = 0* | *Do no use egress flow field, and agents find their ways by none-flow algorithms. In this solver agents may go to their destination site if no exit is specified in the input file. Currently, this solver has some problems to be solved. So users have no direct access to this solver by using GUI. In brief if no exit is specified in the input file, Solver=0 will be enabled automatically because egress flow field cannot be computed with no exit.* |

*Agent-based model (ABM) describes interactions among individual agents and their surroundings, and four types of entities are illustrated in Figure 1, which are agents, walls, doors and exits. The red lines are walls and green and red rectangular areas are doors and exits, respectively. The round disks are agents in social interactions and movement. A discrete mesh field is also generated based on the position and shape of walls, doors and exits, where red dots in Figure 1 are open regions that agents can go through, and the blue crosses set up the boundary regions in the mesh. The simulation object is created by reading in such entities from the input files as shown in Figure 1. The simulation object will be configured and computational work are executed to dump output data, which can be visualized by 2d viewer in our program. In the following sections we will sequentially introduce these topics.*

*The rest of this manual is organized as below. In Section 2 we present how to prepare input data for agents, walls, doors and exits. Such data are used to initialized a simulation object. The theoretical model is briefed in Section3, and the configuration and visualization of the simulation object is introduced in Section 4. Analysis and visualization of output data are mainly discussed in Section 5, and we also briefly introduce how to write a plain-text configuration file in Section 6.*

## *2. Preparing Input Data for Simulation*

*In crowdEgress simulation platform options are provided to users to either use existing FDS input files to create compartment geometries or specify them in a csv input file. In current version only one-floor crowd simulation is supported. So if there are multiple evacuation meshes in FDS input files, they should all belong to the same z interval in the vertical direction (z axis). By using FDS input files the walls are created by &OBST, and the doors are specified by &HOLE or &DOOR. The exits are obtained from &EXIT in FDS input files. In our simulation program FDS data are read by the following functions in data.py, which are readOBST(), readPATH() and readEXIT(). If users want to find more about how FDS define a compartment area, please refer to FDS UserGuide for more information [McGrattan et. al., 2025].*

*If users do not use FDS input files, the above entities can be alternatively specified by using a csv file as introduced below, and users need to write data blocks in csv file to define walls, doors, exits and agents, and such data blocks are claimed by &Wall, &Door, &Exit and &Agent as the first element in the data block.*

### ***2.1 Preparing Data Blocks in CSV Input File***

***Walls**: Walls are obstructions in a compartment geometry that confine agent movement, and they set up the boundary of a room or certain space that agents cannot go through. &Wall is the identifier for the data block of walls. In other words, &Wall claims that this data block describes walls in the simulation, and thus &Wall should be written as the first element in the data block, namely, the most upper left element in the data block. Please refer to examples for more details.*

*In our program walls are either in line or rectangular shape. If any users are interested, please feel free to extend the wall types to circular or poly-angular shapes. If users import walls from a FDS input file, the walls are created as a rectangular type and it corresponds to &OBST in FDS input file. If users specify a line obstruction, it is expected to input the position of starting point and ending point of a line. If users specify a rectangular obstruction, it is expected to input the diagonal position of two points to shape a rectangular area. The specific features of walls are listed in sequence as below.*

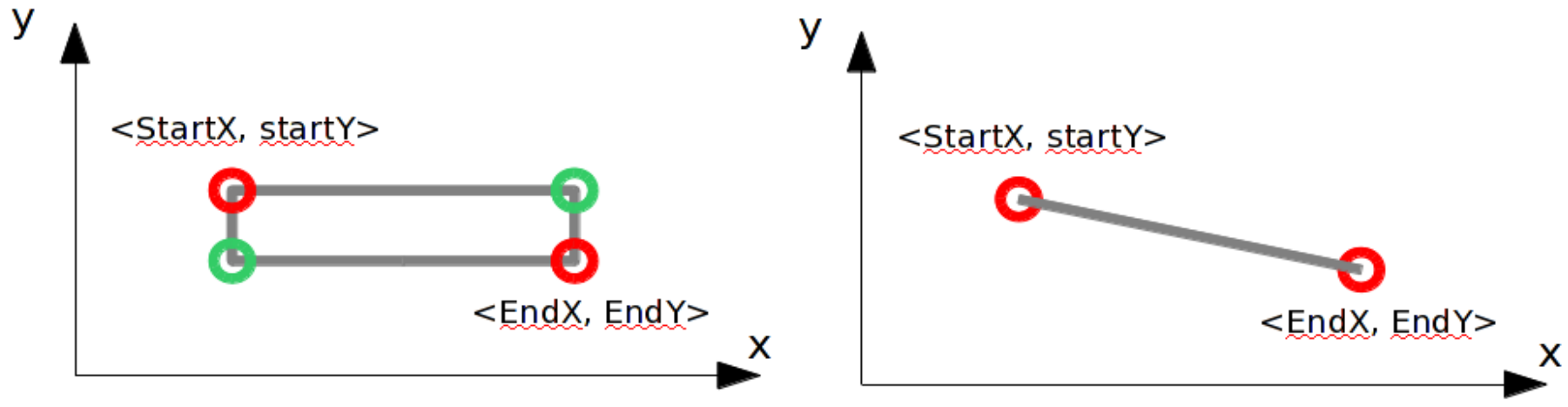

*(a) Rectangular Type* *(b) Line Type*

***Figure 2. Create Walls in Rectangular Type or Line Type.***

*<name>: Name assigned to the wall, and it is arbitrarily given by users. Leave it blank if no name is assigned.*

*<startX, startY>: One diagonal point for rectangular obstruction; Or starting point for line obstruction.*

*<endX, endY>: The other diagonal point for rectangular obstruction; Or ending point for line obstruction.*

*<direction>: Direction assigned to the obstruction so that agents will be guided when seeing this obstruction, especially when they do not have any target door or exit. The direction means if the obstruction provides agents with any egress information (e.g., exit signs on the walls which direct agents to the location of exits). The value could be +1 for positive x direction, -1 for negative x direction, +2 for positive y direction and -2 for negative y direction. If no direction is given, the value is 0 by default, which means the obstruction does not provide any information for egress. User may optionally leave it blank such that the default value 0 is used.*

*<shape>: Walls could be given either in rectangular shape or line shape as shown in Figure 2. The default shape is 'rect', which means the rectangular shape. Use string 'line' if the obstruction is specified in line shape.*

***Table 2. Data Block of Wall***

| &Wall | 1/startX | 2/startY | 3/endX | 4/endY | 5/direction | 6/shape |
|---|---|---|---|---|---|---|
| Wall Down | 0 | 0 | 10 | 0.5 | 0 | rect |
| Wall Top | 0 | 9.5 | 10 | 10 | 0 | rect |
| WallLeft | 0 | 0 | 0.5 | 10 | 0 | rect |
| WallRight | 9.5 | 0 | 10 | 10 | 0 | rect |

***<u>Door and Exit</u>:*** *Doors are passageways that direct agents toward certain areas, and they may be placed over a wall so that agents can get through the wall by the door. Doors can also be placed as a waypoint if not attached to any walls, and they can be considered as arrows or markers on the ground that guide agent egress movement. In brief doors affect agent way-finding activities and they are useful for agents to form a roadmap to exits. In current program doors are only specified as rectangular object, and the data block of doors is claimed by &Door, which should be written as the first element in the data block. Please refer to examples for more details.*

*Exits are a special types of doors which represent paths to the safety. Thus they may be deemed as final path to reach safety areas, and computation of an agent is complete when the agent reaches an exit. An exit is usually placed over a wall like doors, but it can also be put anywhere independently without walls. In the program exits are only defined as rectangular areas. The data block of exits are claimed by &Exit, which should be written as the first element in the data block. Please refer to examples for more details. The specific features of doors and exits are the same, and are listed in sequence as below.*

*<name>: Name assigned to the door/exit, and it is arbitrarily given by users and optionally visualized in the simulation window (pygame screen). Users could identify each door or exit by their names assigned in the input file. If no name is assigned, please leave the cell blank in the input csv file.*

*<startX, startY>: One diagonal point for rectangular door/ exit.*

*<endX, endY>: The other diagonal point for rectangular door/exit.*

*<direction>: Direction assigned to the door or exit so that agents will be guided when seeing this entity, especially when they do not have any target door or exit. The direction implies if the door or exit provides evacuees with any egress information such as exit signs or not. The value could be +1 for positive x direction, -1 for negative x direction, +2 for positive y direction and -2 for negative y direction. If no direction is given, the value is 0 by default, which means the door/exit does not provide any information of egress or exit locations. The direction is partly used in the way-finding algorithm when agents adapt their routes to exits. If users do not know how to specify the direction, simply leave it blank such that the default value 0 is used. In addition if users select solver=1 or 2 for simulation, the egress flow field will be calculated and it will automatically assign the direction for each door, and the direction value in the csv file will be overwritten.*

*<shape>: All the doors/exits are only in rectangular shape. The default value is 'rect', which means the rectangular shape of doors or exits. User could leave it blank to use the default value.*

***Table 3. Data Block of Exit***

| &Exit | 1/startX | 2/startY | 3/endX | 4/endY | 5/direction | 6/shape |
|---|---|---|---|---|---|---|
| Exit Down | 4.5 | −0.3 | 5.5 | 0.3 | 0 | rect |
| Exit Top | 4.5 | 9.7 | 5.5 | 10.3 | 0 | rect |
| Exit Right | 9.3 | 4.3 | 10.3 | 5.9 | 0 | rect |

***Table 4. Data Block of Door***

| &Door | 1/startX | 2/startY | 3/endX | 4/endY | 5/direction | 6/shape |
|---|---|---|---|---|---|---|
| Door Down | 4.49 | 0 | 5.51 | 0.6 | 0 | rect |
| Door Top | 4.49 | 9.4 | 5.51 | 10.1 | 0 | rect |

***<u>Agents</u>**: Finally and most importantly, agents are the core entity in computation process. They interact with each other to form collective behavior of crowd. They also interact with above types of entities to form egress motion toward exits. The resulting program is essentially a multi-agent simulation of pedestrian crowd. Each agent is modeled by extending the well-known social force model. The model is further advanced by integrating several features including pre-movement behavior, group behavior, way-finding behavior and so forth. In current program the data block of agents are claimed by &Ped or &Agent, and they are written as the first element in the data block.*

*<name>: Name assigned to each agent and it is arbitrarily given by users and optionally visualized in the simulation window (pygame screen). The names of agents are given in the first column of the data block. Users may leave this column blank if no names are used.*

*<InitalX, InitialY>: Initial position of an agent in 2D planar space. Please note that such initial position is required to be given by users and no default values are used in program.*

*<InitialVx, InitialVy>: Initial velocity of an agent in 2D planar space. The default value is <0.0, 0.0>, which means that agents have zero velocity in the initial state, i.e., they are standing still in the initial position at time point t=0.0.*

*<tau>: Tau parameter in the social force model, or as usually called relaxation time in many-particle systems or statistical physics, and it critically affects how fast the actual velocity converges to the desired velocity. The default value is 0.6.*

*<tpre>: Time period for pre-movement stage. Within this time period agents do not select and move towards an exit. The default value is 10 seconds. The timeline model of pre-movement phase and movement phase is illustrated in Figure 3. In the pre-movement phase evacuees are inclined to collect information rather than actually move to*

*any exit. The pre-movement behavior is critically studied in this simulation platform, where existing models in opinion dynamics are applied to aggregate pre-movement time of many individuals. As shown in Figure 3 the simulation starts at t=0, namely the time of alarm, and all the agents receive the alarm information immediately at t=0. Let* $tpre_i$ *denote the pre-*movement *time interval of agent i, and it is updated by mixing itself with a weighted average of others given certain social relationship among individuals. When the simulation t exceeds* $tpre_i$*, agent i starts to move to an exit selected.*

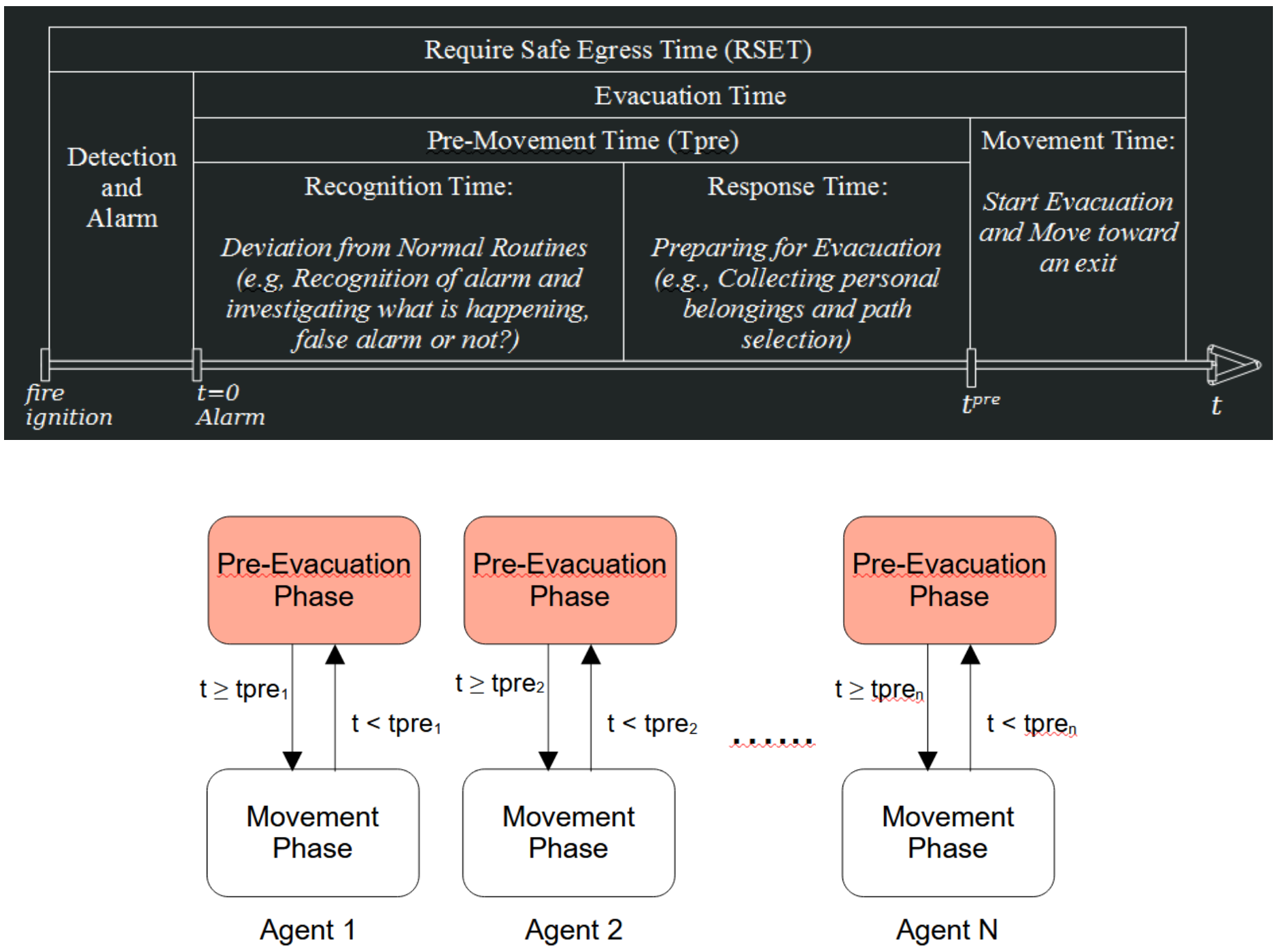

***Figure 3. Time Line Model for Pre-Movement Behavior.***

*<p>: Parameter p in opinion dynamics, and it indicates how an agent's opinion/decision is impacted by surrounding others, and it critically affects herding effect in collective behavior. The measurement of this parameter is within [0, 1], and an agent's opinion/decision completely follow others if p=1. In contrast, if the agent makes decision only based on his or her own opinion, the parameter p=0, and this usually implies that the agent is a leader in a group. Thus, we also call p by decision balance parameter in this manual because it determines how an individual agent keeps balance between his or her own opinion and others' opinion in decision making process. In sum our model and algorithm critically consider the social topology of many individuals in exit selection behavior. In a sense it is useful to describe leader-and-follower relationship, or more generally as called the herding effect in collective behavior. This is especially evident when people are responding to an emergency (Low, 2000), where excessive time-pressure weakens the ability of logical thinking and reasoning, and independent decision making becomes more and more difficult. Thus, people are more inclined to follow others (e.g., surrounding others' decision or a leader's decision) rather than make decisions by themselves. In other words the choice of others or a leader in social group are very important, and it may significantly impacts an agent's probability distribution of exit selection when solver=2. In our algorithm parameter <p> is the weight for an agent to timely integrate such new information into one's decision-making process.*

*<pMode>: This parameter indicates whether or how parameter p is dynamically updated. Currently there are three methods to be selected: random, fixed or stress. If 'random' method is used, it means that parameter p is randomly generated in the interval of [0, 1] by uniform distribution. If 'fixed' method is selected, parameter p is given by the initial value in the input csv file, and it keeps constants through the simulation. The 'stress' method will adapt parameter p by each agent's stress level. In principle the value of p increases if an agent feels more stressed. This method is being tested and it involves a computational model to adapt parameter p dynamically to surroundings.*

*<p2>: The egress layout and facilities determine a quantitative utility of each exit, and thus random utility model is useful to timely integrate such information into one's decision making process (Lovreglio, Fonzone and dell'Olio, 2016). For example, the utility of an exit is larger for an agent if it is closer to the agent's location, and thus it is more likely for the agent to select it. Extensive research has been conducted on how to properly develop such a utility function, and it mainly refers to social and economic models and theories. In our algorithm parameter <p2>*

*is the weight for an agent to timely integrate such new information into one's decision-making process. This parameter affects how much an agent tends to make a decision based on the information timely collected from the surrounding facilities such as the distance to the target exits or received guidance from broadcast. The measurement of this parameter is within [0, 1], and an agent's opinion/decision completely follows the received information if p2=1. In contrast, if the agent makes decision only based on his or her past experience and the new information is completely ignored, the parameter p2=0.*

*To better understand how <p> <pMode> and <p2> works in our simulation model and algorithm we will take the exit selection for example, and an agent's selection of an exit is a stochastic process which involves three basic conventions: memory of past experience, bounded rationality, and herding instinct. Thus, for each individual agent the probability is aggregated by three major factors: (1) past experience or prior habit of using certain exit. (2) Collecting information and reach a decision on current situation (3) Hearing and learning from other individuals. Now it is feasible to build up a probabilistic model to link different factors in a statistical sense to update the exit-selection probability for sovler=2 (Wang et. al., 2008; Wang et. al., 2022).*

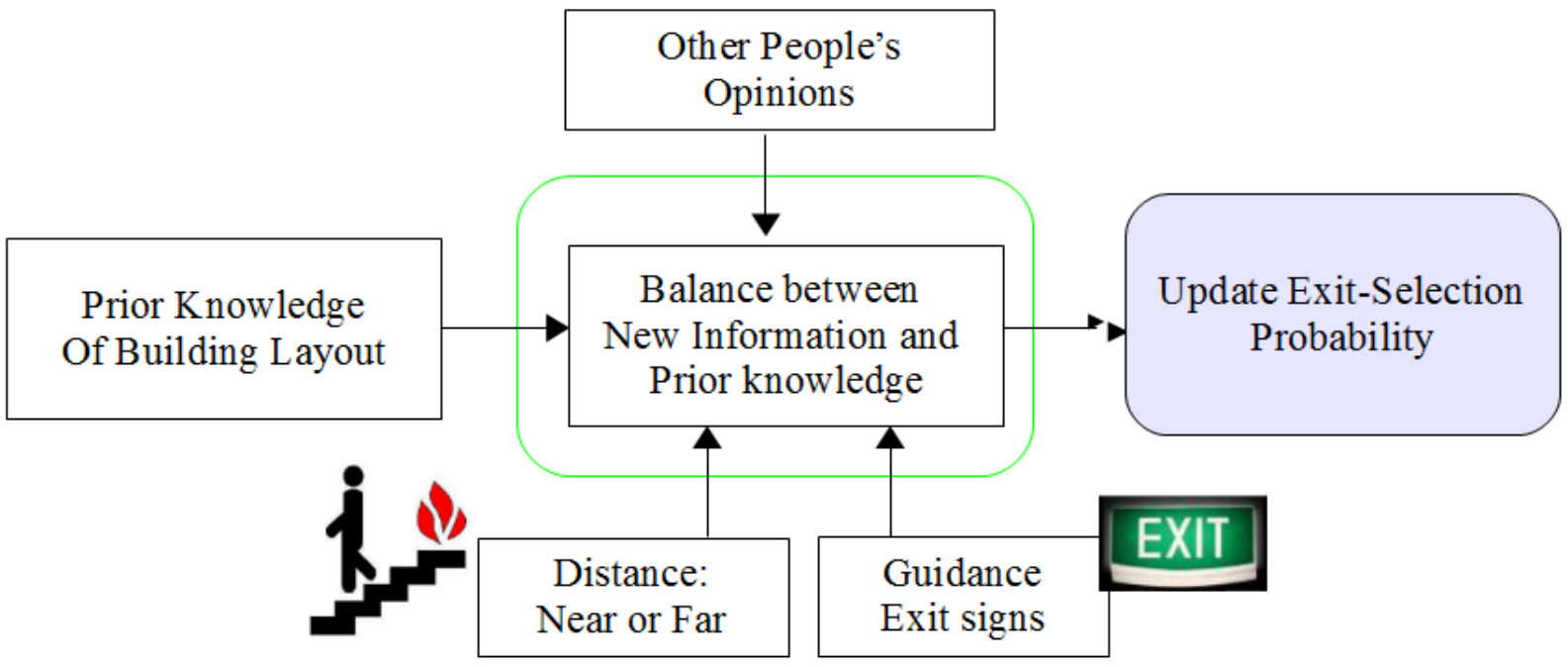

***Figure 4. Information fusion for timely updating the exit-selection probability.***

*<Range>: The range to determine when agents have opinion exchange. Such interaction could also be understood as herding effect or group opinion dynamics, which means agents exchange opinions by talking. This parameter is similar to the interaction range in Vicsek flocking model (Vicsek et. al., 1995), where two agents interacts if their physical distance is within this range parameter. In our simulation model this range parameter mainly determine the attention list and talking list for each agents.*

*<moveMode>: The type of way-finding behaviors. Some agents may actively search for exits while others may just follow the crowd. In current simulation all agents either follow the egress flow field or find their ways based on solver selected, and thus this parameter is not actually used in existing version of code.*

*<inComp>: a boolean variable to indicate if the agent is in computation loop or not. Normally it is given true. If users want to quickly remove an agent in computation loop, please assign it false for convenience.*

*<tpreR>: The type of arousal level in the pre-movement phase. Some agents may actively exchange information with others and they are assigned with the highest level 3 in simulation. The level 2 of arousal implies that agents will passively interact with others, and they only receive information but not actively search for new information from surroundings. The level 1 is the lowest arousal level in our model, meaning that agents stay static in their initial positions, possibly consider the alarm as a false alarm or not aware of the alarm (e.g., in sleep). The arousal level will increase irreversibly in the simulation process.*

*<talkTau>: Tau parameter when agents have opinion exchange, and it affects how fast the actual velocity converges to the desired velocity when agents are in states of talking. The default value is 0.6.*

*<talkProb>: The probability to determine if agents have opinion exchange. Such interaction could also be understood as herding effect or agents exchange opinions by talking.*

*<mass>: The mass of agents.*

*<radius>: The radius of agents.*

*<tau_tpre>: The relaxation time when an agent is in pre-movement stage.*

*<tau_talk>: The relaxation time when an agent exchanges opinion by talking.*

*<DestX, DestY>: Destination position in 2D planar space. This value is almost not used in current computational loop because the destination position is automatically determined by the exit selection algorithm. When the exit is selected by an agent, the destination position is given by the exit position.*

***Table 5. Data Block of Agents***

| &Agent | 01/ IniX | 02/ IniY | 03/ IniVx | 04/ IniVy | 05/ tau | 06/ tpre | 07/ p | 08/ pMode | 09/ p2 | 10/ Range | 11/ aType | 12/ inComp | 13/ tpreR | ... |
|---|---|---|---|---|---|---|---|---|---|---|---|---|---|---|
| Agent0 | 6.3 | 2 | 0.0 | 0.0 | 1.3 | 18 | 0 | fixed | 0.7 | 10 | active | 1 | 10 | ... |
| Tuck | 7.3 | 3 | 0.6 | 0.2 | 0.6 | 7 | 0.6 | random | 0.6 | 5 | active | 1 | 20 | ... |
| James | 7.3 | 6 | 0.3 | 0.7 | 0.6 | 6 | 0.6 | stress | 0.6 | 6 | search | 1 | 16 | ... |

## 2.2 Important Tips:

*(1) The sequence of &Wall &Door &Exit &Agent could be arbitrarily changed within an input file. Users may either first specify walls, or doors or agents. However, data feature inside a data array could not be changed in sequence. For example, users must first give <startX>, <startY>, <endX>, <endY>, and then specify the direction and shape if needed. Users cannot change the sequence to be <startX>, <endX>, <startY>, <endY>, or specify the direction or shape before them. All the features for walls, doors, exits and agents are read in certain order, and such order cannot be altered in current version (version 2.3). New contributors are welcome to improve this functional setting in data_func.py such that this sequence can also be adjusted by users (e.g., using the DataFame in python pandas module). Please refer to readWalls(), readDoors(), readExits() and readAgents() in data_func.py for more details.*

*(2) As above we exemplify all the input data by table-like data array, and users can select a table processing software such as Microsoft Excel or GNUMeric to show such data. However, please remember that csv (comma separated values) is the basic format of the above data file, namely all the data cell in the input file are separated by comma. Those comma are omitted in Excel or GNUMeric, but are shown in any text editor. Please note that when users save the data in such table processing softwares, the csv data could be slightly extended automatically, for example, adding many empty cell at the end of each line in the data array. Usually such empty cells are useless and our program still works well with them. If not, users should check the text form of csv files without using any table-processing softwares.*

*(3) In the latest version of the program we also integrate an easy text-based csv editor in the GUI, where tab (/t) is used to align the data element in csv file. Our text-based GUI avoid adding useless empty cells at the end of each line. There is also a treeview-based csv editor integrated in the GUI, and users can use it to modify the input csv data and also no empty cells will be added at the end of each line. The treeview-based csv editor is shown in Figure 23. For more information on using the above csv editors in GUI, please refer to Section 4 for more details.*

*(4) Please note that all the entities including walls, doors, exits and agents are also assigned with a unique ID number, and this number is automatically given by the program. The ID number starts from zero and it increases based on the sequences of entities specified in the input csv file. Both of the name and ID numbers could be visualized in simulation window (pygame screen). If users hope to change this ID number, the only method is adjusting the sequence when claiming these entities in the data array of input csv file. For more details, please refer to the python scripts: the objects of walls, exits and doors are programmed in obst.py in this project, and the object of agents is programmed in agent.py. Users are welcome to review the source code in this project.*

*(5) If users start to specify any entities in the data block of walls, doors, exits or agents, the shade areas in Table 2 – 5 are the required parameters that users must input to initialize a simulation object. If not, the simulation object cannot be initialized properly. For other parameters, if users do not understand what they exactly means, please omit them in the csv input file and the default value will be automatically chosen by the program. Please note that the data block for agents must be included in the input csv file. In contrast, the data blocks for wall, doors, and exits can be omitted (Do not claim the data blocks for walls, doors or exits, or claim empty data blocks for them with no entities included.*

*(6) The multi-compartment layout can be created by either FDS input file or the data blocks in our csv input file. Besides, there is also third method to create/modify a multi-compartment layout, i.e., drawing the layout manually in the section of TestGeom. In other words, users can completely omit the specification of the multi-compartment layout in csv input file, and launch the program. By default, the program will first go to the section of TestGeom (Testing of Geometric Layout) as shown in Figure 9. In this section, users are able to manually draw the walls, doors or exits on the screen, and directly check if the geometric layout is satisfied or not. If not, users can press <z> in*

*keyboard to cancel their latest drawing on the screen. If a FDS input file or a csv input is provided, the section of TestGeom will automatically show the layout in the input file, and users can add or delete any walls, doors or exits on the screen. This graphical method only enables users to easily create or modify the walls, doors, and exits, but not the agent. Agent features must be well specified in the input csv file, and only the initial position of agents can be manually changed in TestGeom. For more features on TestGeom, please refer to Section 4.2.*

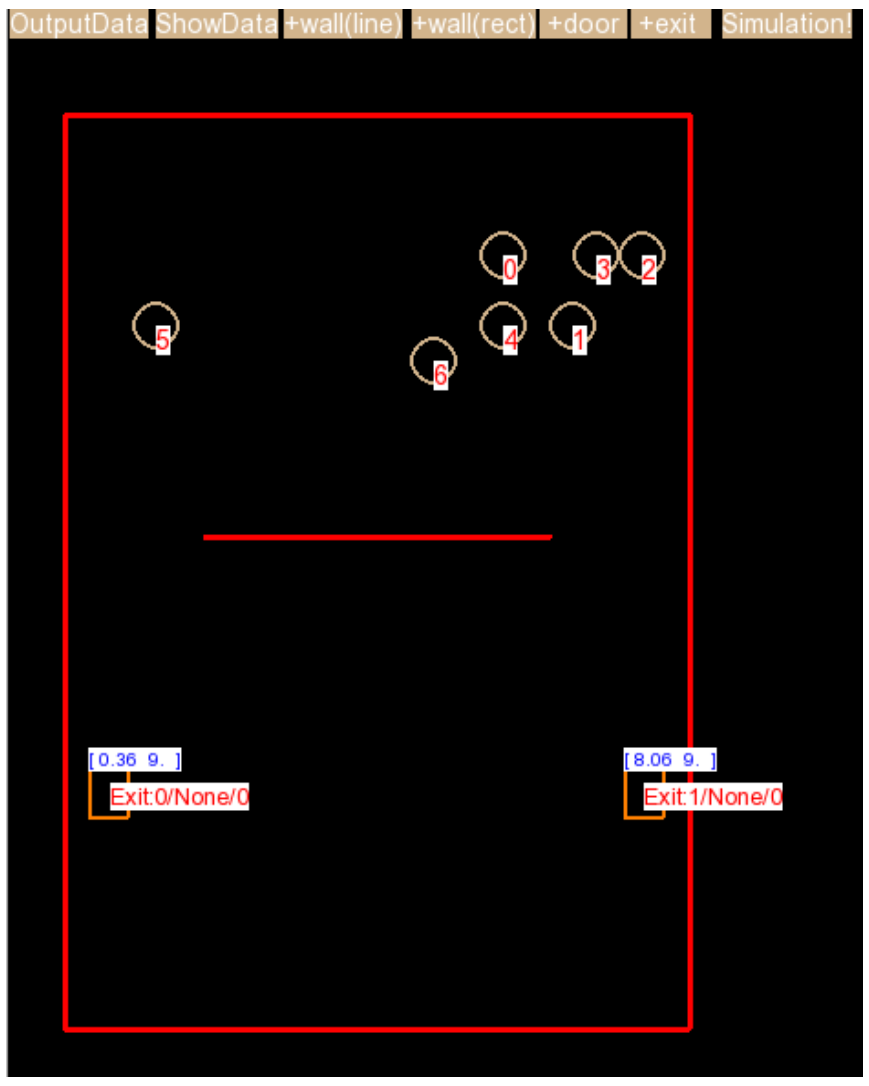

***(a) TestGeom***

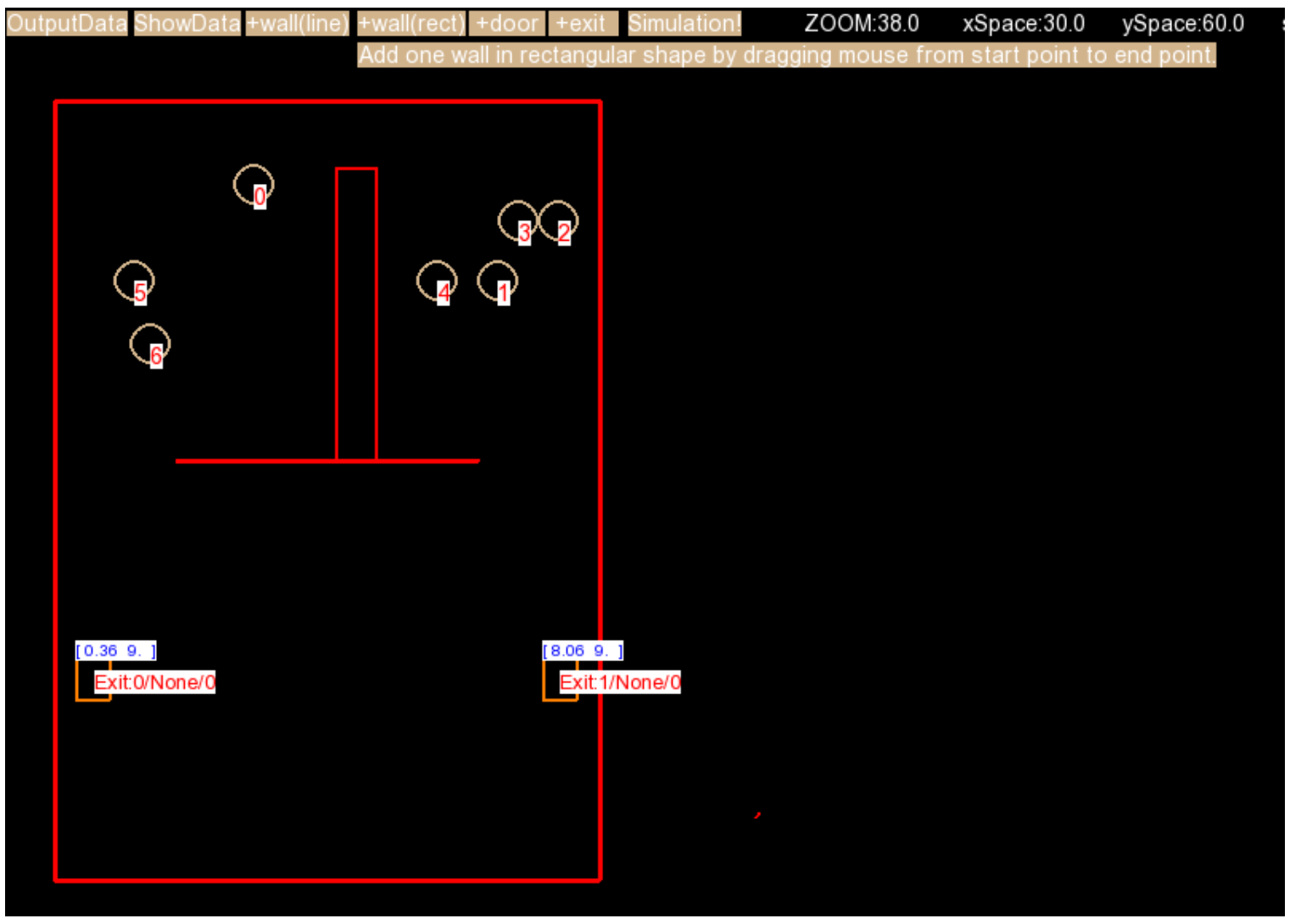

***(b) Add walls or move agents in TestGeom.***

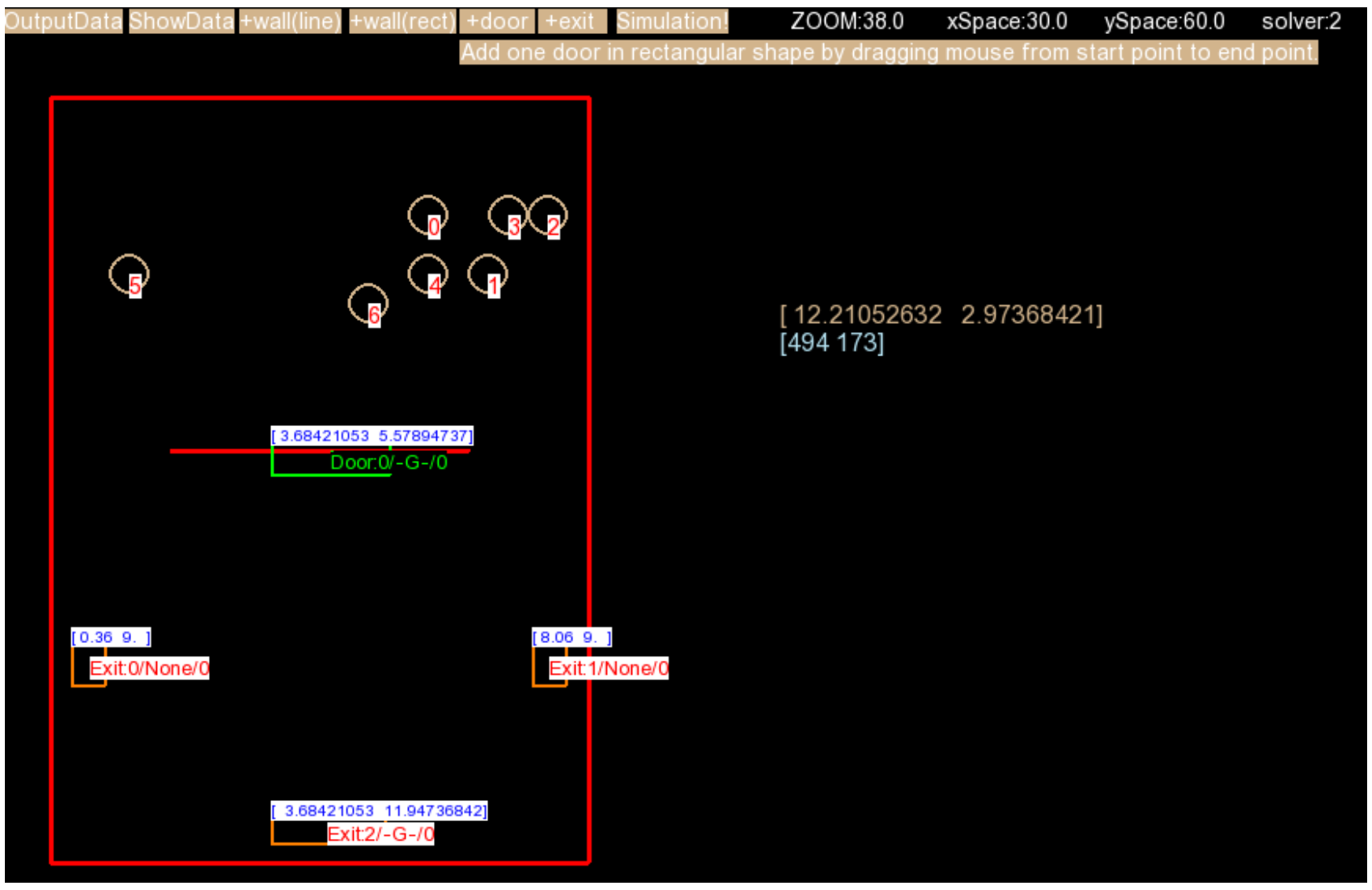

***(c) Add doors or exits in TestGeom.***

***Figure 5. Initialize Simulation Object in Pygame Screen: Adding Doors or Exits.***

*The above model parameters are related to the theoretical model of agents in various scenarios such as in pre-movement stage or in talk state. In brief these parameters significantly affect how an agent interacts with others and how they decide the egress routes in collective behavior. Please refer to our article (Wang et. al., 2022) for more details of the theoretical model. As below we only highlight the key features of our model as shown in Figure 6.*

## *3. Brief Introduction of Theoretical Model*

*In current version of program (version 2.3) we take into account the environmental stressors including egress facilities (e.g., alarm, guidance), and evacuees respond to such stressors and move to the safety. In a sense such stress is caused by mismatch between psychological demand and realistic situation (Staal, 2004), and we summarize the mismatch in terms of velocity and interpersonal distance as shown in Figure 6: the psychological demands are abstracted as desired velocity* $\mathbf{v}_i^0$ *and desired interpersonal distance* $d_{ij}^0$ *while the physical reality is described by the physical velocity* $\mathbf{v}_i$ *and distance* $d_{ij}$*. The difference of two variables measures how much people feel stressed, by which people are motivated into certain behavior. The motivation is abstracted as the driving force and social force (Wang, 2016; Wang and Wang, 2021), and the entire process formalizes the stimuli-reaction model (S-R model) in psychological view of behaviorism.*

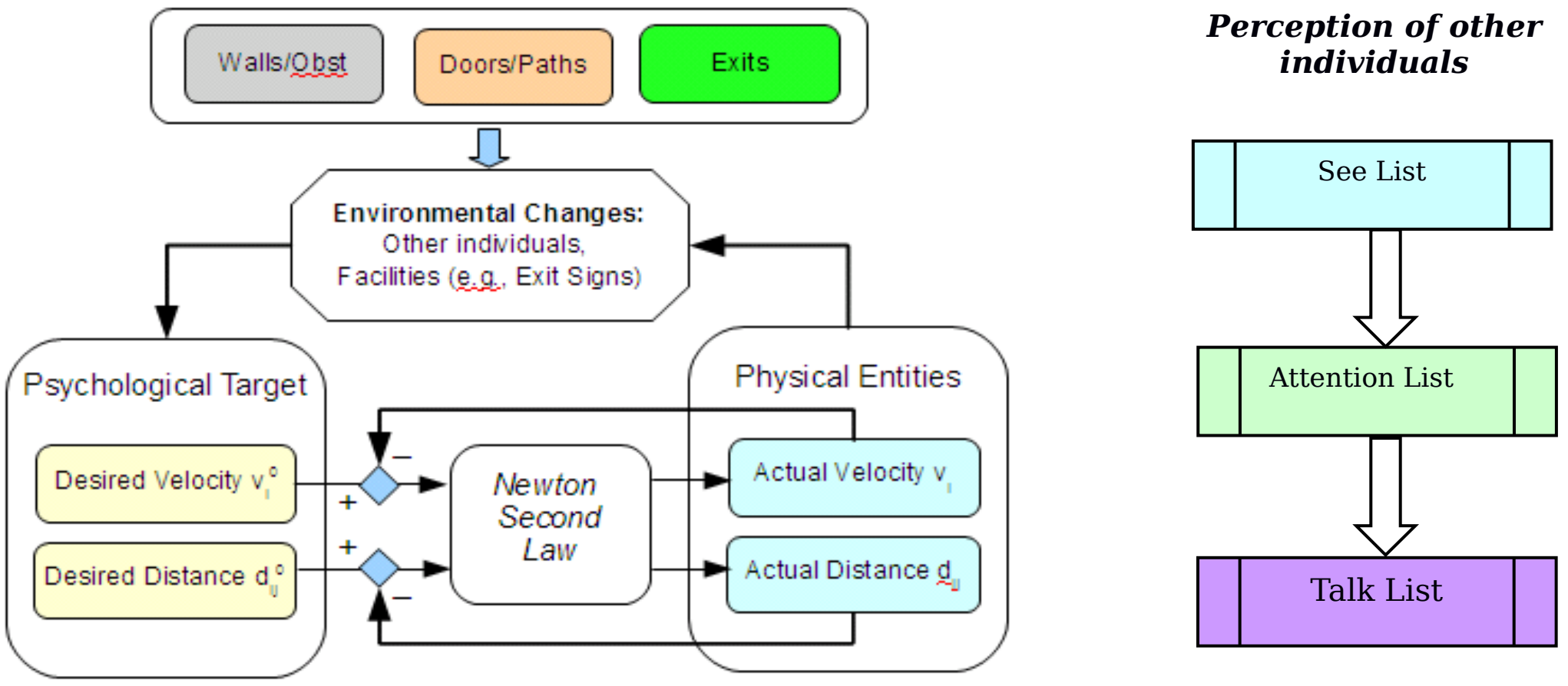

***Figure 6. Agent-based model with Environmental Facilities.***

### ***3.1 Forced-Based Interactions:***

*Commonly, emergency egress is a stressful situation. When people hear the fire alarm,, they respond to such environmental stressors and begin to escape to the safety (Sidiropoulos, Kiourt, Moussiades, 2020, Templeton et. al., 2024). Each individual agent's behavior is described by physics laws as given below.*

$$m_i(d\mathbf{v}_i/dt) = \boldsymbol{f}_i^{drv} + \sum_j \boldsymbol{f}_{ij} + \sum_w \boldsymbol{f}_{iw}$$

*This approach is widely called a force-based model, where the mass of an individual is denoted by* $m_i$*, and the change of the individual's instantaneous velocity* $\mathbf{v}_i(t)$ *is determined by the total force. The forces motivates people to adapt to surrounding stimuli, and they consist of driving force* $\boldsymbol{f}_i^{drv}$*, interaction force* $\boldsymbol{f}_{ij}$ *as well as boundary wall force* $\boldsymbol{f}_{iw}$*. The driving force functions as an energy source that drives people to move to a destination exit, and it is commonly specified by* $\boldsymbol{f}^{drv} = \boldsymbol{F}^{drv}(\mathbf{v}_i^0 - \mathbf{v}_i)$*, where* $\boldsymbol{F}^{drv}(.)$*is generally a monotonically increasing function. The interaction force* $\boldsymbol{f}_{ij}$ *describes the social-psychological tendency of two individuals to keep proper distance (as called the social-force), and if people have physical contact with each other, physical forces are also taken into account, i.e.,* $\boldsymbol{f}_{ij} = \boldsymbol{f}_{ij}^{soc} + \boldsymbol{f}_{ij}^{phy}$*. In this article we mainly focus on the social force term. In Helbing et. al., 2002 and 2005 this term is given in an exponential form, and we will integrate a new concept of desired distance* $d_{ij}^0$*into this force term. In a general sense we specify this force term by* $\boldsymbol{f}_{ij}^{soc} = \boldsymbol{F}^{soc}(d_{ij}^0 - d_{ij})$*. The interaction of an individual with obstacles like walls is treated analogously, and denoted by* $\boldsymbol{f}_{iw} = \boldsymbol{f}_{iw}^{soc} + \boldsymbol{f}_{iw}^{phy}$*, and this force specifies boundary of people's motion. When an agent is sufficiently close to a door or exit,* $\boldsymbol{f}_{iw}^{soc}$ *also include attractions calculated between the agent and the door or exit, and such attractions are called door force in the output text file, and it is mainly useful for our test of Solver=0. As for Solver=1 or 2, because agents are guided by egress flow fields, it is not that useful to calculate such door forces.*

*In our agent-based model we will mainly focus on* $\boldsymbol{f}_i^{drv}$ *and* $\boldsymbol{f}_{ij}^{soc}$ *in the background of evacuation study. In addition agents can choose information used for the representation of social environments and are able to cast their attention purposely on certain details. The feature of selective attention is critically modeled by three lists in our simulation platform, which are see list, attention list and talk list. For simplicity, users could understand such three lists as three different sets, and the talk list is a subset of the attention list, and the attention list is a subset of see list. In other words, we assume that an agent first sees surrounding others such that all the agents in his or her viewsight are included in the see list. The member in the see list will be selectively added in the attention list if the* *agent will cast his or her attention to the member in the see list, and such selection process is calculated based on the social topology among agents. We will introduce this issue soon later. Finally, the agent may talk to some members in the attention list with certain probability, and such talking behavior forms the talk list for the agent. In practical computing the lists are represented by matrix-based data, and updated in the simulation process. For example, suppose there are n agents in the computation loop, and we have boolean matrices as SEE=*$[see_{ij}]_{nxn}$*, ATT=*$[att_{ij}]_{nxn}$ *and TALK=*$[talk_{ij}]_{nxn}$. *The element of* $see_{ij}$ *=1 means that agent i is able to see agent j, and* $att_{ij}$*=1 means that agent i pays attention to agent j, and* $talk_{ij}$ *=1 means that agent i is talking to agent j. Obviously, we have* $see_{ij}$ *=1 if* $att_{ij}$ *=1, and similarly have* $att_{ij}$ *=1 if* $talk_{ij}$ *=1. These boolean matrices are updated through the simulation process, and they critically determine how an individual agent selects surrounding others for social interactions. The three lists are declared as class variables in our program. Different from the instance variables, such class variables are declared inside of a class, and shared across all instances (objects) of the class. Thus, the three lists belong to the class of agent, but not any specific instance of agent class. The diagonal elements in the three lists are all ones. Other elements in the three lists are updated by three functions in agent.py, which are named by updateSeeList(), updateAttentionList(), updateTalkList(), and such functions are used to update the class variables declared at the beginning of agent class.*

*Next we will emphasize that* $\boldsymbol{f}_i^{drv}$ $\boldsymbol{f}_{ij}^{soc}$ *and* $\boldsymbol{f}_{iw}^{soc}$ *are all subjective entities coming from people's opinions, and they are generated intentionally by people through foot-ground friction on physics basis. These forces essentially describe how an individual perceive and react to the outside environmental stimuli. In particular* $\boldsymbol{v}_i^0$ *and* $d_{ij}^0$ *are non-physics entities for they exist in people's opinion, not in the physical world, and we have well explained the psychological background of the model in Wang, 2016, and critically modified the key concept of social force in consistency of both physical laws and psychological principles.*

*Opinion dynamic model will be jointly used with the above force-based model to investigate how agents interact in social context. Users may interpret such opinions by* $\boldsymbol{v}_i^0$ *and* $d_{ij}^0$ *in the driving force and interaction force as above. This means that each agent's* $\boldsymbol{v}_i^0$ *and* $d_{ij}^0$ *interact with surrounding others, and this feature is especially useful to characterize the collective motion of many agents (Viscek, 1995).*

***Table 6. On Conception of Stress and Force-Based Model***

| | *Opinion (Psychological Characteristics)* | *Behavior (Physics-Based Characteristics)* | *Difference between subjective opinion and objective reality* | *Forced-Based Term* |
|---|---|---|---|---|
| ***Time-Related Characteristics*** | **desired velocities** $\boldsymbol{v}_i^0 = v_i^0 \boldsymbol{e}_i^0$ | **actual velocities** $\boldsymbol{v}_i = v_i \boldsymbol{e}_i$ | ***Time-Related Stress: Velocity*** $\boldsymbol{v}_i^0 - \boldsymbol{v}_i$ | ***Driving Force*** $\boldsymbol{f}_i^{drv}=\boldsymbol{F}^{drv}(\boldsymbol{v}_i^0 - \boldsymbol{v}_i)$ |
| ***Space-related Characteristics*** | **desired distance** $d_{ij}^0$ | **actual distance** $d_{ij}$ | ***Space-related Stress: Distance*** $d_{ij}^0 - d_{ij}$ | ***Social Force*** $\boldsymbol{f}_{ij}^{soc} = \boldsymbol{F}^{soc}(d_{ij}^0 - d_{ij})$ |

## ***3.2 Social Groups and Social Topology:***

*The social force* $\boldsymbol{f}_{ij}^{soc}$ *usually includes a significant component which combines individuals into groups. This component is called group force. For example, given agent i and j, the group force excerted on agent i from agent j can be formulated as below.*

$$\mathbf{f}_{ij}^{soc} = \frac{A_{ij}}{B_{ij}}\left(d_{ij}^0 - d_{ij}\right)\exp\left[\frac{\left(d_{ij}^0 - d_{ij}\right)}{B_{ij}}\right]\mathbf{n}_{ij} \qquad \mathbf{f}_i^{drv} = \frac{m_i\left(\mathbf{v}_i^0 - \mathbf{v}_i\right)}{\tau_i}$$

*In the above equation* $A_{ij}$ *and* $B_{ij}$ *are positive constants, which affect the strength and effective range about how two agents are influential to each other and* $\boldsymbol{n}_{ij}$ *is the normalized vector which points from agent j to i. The equilibrium distance* $d_{ij}^0$ *is called desired interpersonal distance, which is the counterpart of desired velocity in the social force model (Helbing and Molnar 1995, Helbing et. al., 2002). The above mathematical description of group force is called group social force because it is extended from the traditional social force to describe the cohesive effect among agents. Other forms of group force can also be applied such as molecular interactions (e.g., Lennard-Jones potential or Morse potential in molecular dynamics).*

*In addition the desired velocity deduces the driving force, which is widely formalized in a linear manner of* $\boldsymbol{f}^{drv} = \boldsymbol{F}^{drv}(\boldsymbol{v}_i^0 - \boldsymbol{v}_i) = m_i (\boldsymbol{v}_i^0 - \boldsymbol{v}_i)/\tau_i$. *Therefore, the group social force and driving force are both functional in a feedback manner to make the physical distance* $d_{ij}$ *or velocity* $\boldsymbol{v}_i$ *approaching towards the equilibrium distance* $d_{ij}^0$ *and velocity* $\boldsymbol{v}_i^0$. *Due to the page limit we will not further describe the mathematical model of social groups in this manual. If users are interested in our social group model, please refer to Wang et. al., 2022 for more details.*

*As opinion dynamics is integrated with the force-based interactions, the problem becomes interesting because each individual opinion and behavior are described as two different features, and such features interact among individuals in a given social topology to shape collective opinion and behavior. In our modeling framework the social topology among individuals is described by a set of matrices, which provide quantitative measures of their social relationship. Such matrices are basically denoted by S, A, B, D in this manual. Matrix S primarily defines the social topology among agents, which is a quantitative measure of social relationship among agents and thus is closely related to the opinion dynamic model. In our simulation platform Matrix S is claimed by &groupS or &groupC as shown in Table 7. Value of S are normalized within the range of [0, 1]. In fact users do not have to input such normalized values of social topology, and the program will automatically normalize such parameters in simulation process.*

***Table 7. Data Block of Social Groups***

| &groupS | Agent0 | Agent1 | Agent2 | Agent3 | Agent4 | Agent5 | Agent6 | Agent7 |
|---|---|---|---|---|---|---|---|---|
| Agent0 | 0 | 1 | 0 | 0 | 0 | 0 | 0 | 0 |
| Agent1 | 0.2 | 0 | 0.5 | 0.3 | 0 | 0 | 0 | 0 |
| Agent2 | 0 | 0.7 | 0 | 0.3 | 0 | 0 | 0 | 0 |
| Agent3 | 0 | 0 | 0.3 | 0 | 0 | 0.1 | 0.6 | 0 |
| Agent4 | 0 | 0 | 0 | 0.2 | 0 | 0.3 | 0 | 0.5 |
| Agent5 | 0 | 0 | 0 | 0 | 0 | 0 | 1 | 0 |
| Agent6 | 0 | 0 | 0 | 0 | 0 | 1 | 0 | 0 |
| Agent7 | 0 | 0 | 0 | 0.5 | 0 | 0.5 | 0 | 0 |

*In the matrix as shown in Table 7 we can generally identify a group which consists of individual 0, 1, 2, 3. In this group individual 0 is completely a follower to individual 1 with a weight of 1.0, and individual 1 also cares about individual 0 with a weight of 0.2. Such valuation represents a kind of leader-and-follower or child-and-parents relationship in social topology. Individual 1 and 2 care about each other significantly (with weight of 0.5 and 0.7), and both of them are also socially bonded with individual 3 (with weight 0.3 and 0.3).*

*In contrast individual 4, 5, 6, 7 are considered as another social group. In particular individual 5 and 6 compose a stable pair because they care about each other mutually (e.g., couples). Individual 7 also follow individual 3 and 5, but individual 3 and 5 do not care individual 7, and thus they are in leader-and-follower pattern.*

*Individual 3 and 4 critically connect the above two groups. However, such connection are directional in a sense that individual 3 and 4 are widely impacted by individual 5, 6, 7 (second group) and individual 0, 1, 2 (first group) are moderately impacted by individual 3 and 4.*

*In this social topology individual 5 and 6 have the leadership, and they directly impact individual 3, 4 and 7, and also mainly rely on individual 3 and 4 to indirectly affect individual 0, 1, 2. In addition individual 3 is the critical bridge who connects these two groups. If individual 3 moves out of this crowd, the two groups become isolated completely.*

*Matrices A, B, D are related to group social force that characterize force-based social interaction among agents, and they can be automatically deduced from the above matrix S to compute the social interaction among agents. For example, a typical rule can be used to compute all the element in A, B, D, such as* $A_{ij}=100s_{ij}$, $B_{ij}=0.001+10s_{ij}$, *and* $d_{ij}^0=1.0$ *if* $s_{ij}\neq 0$. *If* $s_{ij}=0$, $A_{ij}=0$ *and the group force is simply zero. All in all, this issue is a little complicated in mathematics, and we will not talk about this issue in detail in this manual. If readers are interested, please read our academic papers or the source code of the CrowdEgress program.*

### ***3.3 Exit Selection Probability:***

*Way finding refers to how individual agents orientate themselves towards exits, especially within a multi-compartment layout. Several factors influence their decision making in way-finding activities such as guidance information (e.g., exit signs) or other individuals' choice, and it characterizes how agents perceive and integrate various information acquired from surrounding facilities and individuals (Tong and Bode, 2022). In our simulation platform agent i starts to move towards an exit when the simulation time* $t>tpre_i$. *The problem could be formalized in two steps: (a) How to choose a destination exit; (b) How to select a proper route to reach the exit.*

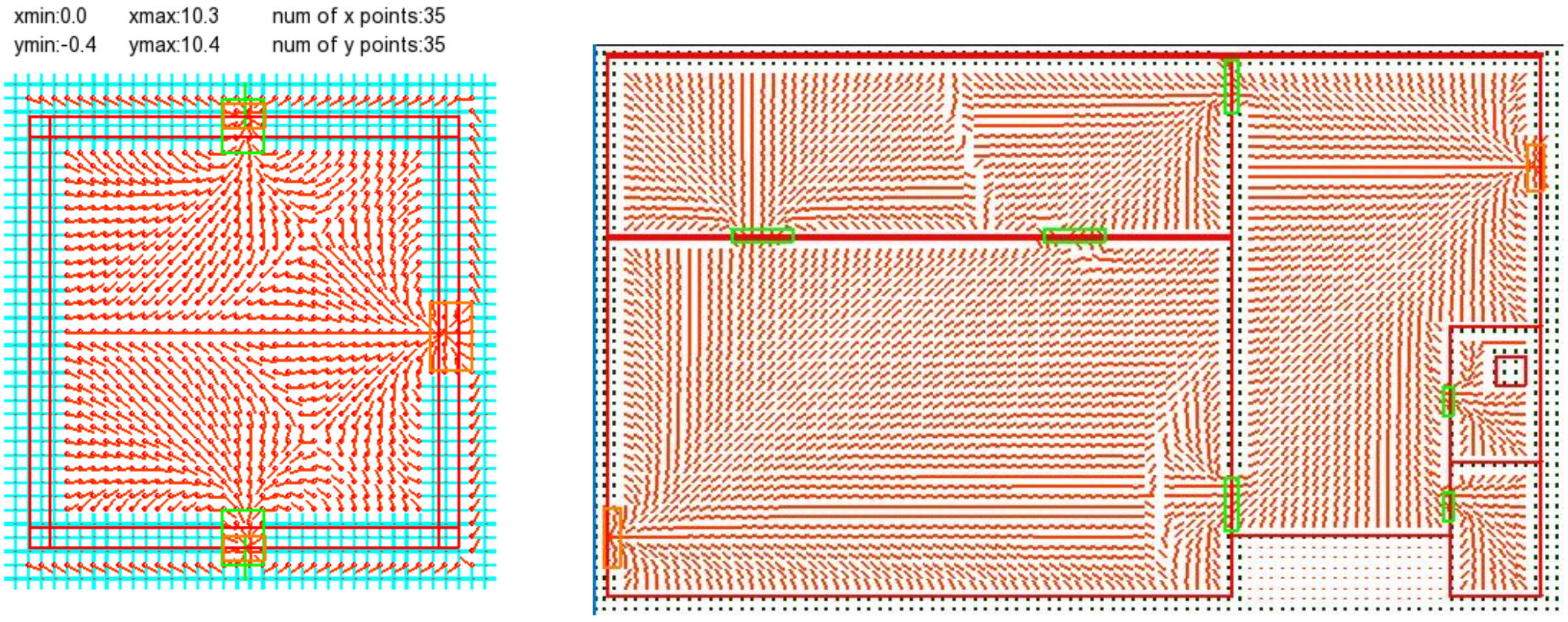

*(a)* *(b)*

***Figure 7. Flow field of using the nearest exit.***

*If users select the default setup of using the nearest-exit strategy, it implies that each agent will select the exit nearest to their current locations, and no exit-selection probability is involved. This means that solver 1 is used in our program as listed in Table 7, and the program calculates an egress flow field for the nearest-exit strategy, and agents follow this field to move towards the nearest-exit to their locations. The egress flow field for nearest-exit strategy is illustrated in Figure 7.*

*Another solver deals with a more complicated scenario, where each individual's choice is determined by exit-selection probability distribution. This means that solver=2 is selected in our program as shown in Table 1. In other words we suppose that there are several candidate exits known by evacuees and we simulate how they select one among the known exits and choose a proper way to reach them.*

*The exit probability is initialized by a data array as exemplified in Table 6, where each row corresponds to each agent and the column refers to exit selection probability. The entire data array is claimed by &Agent2Exit or &Ped2Exit, which is written in the position of the first element. All the agents select exits based on the discrete probability distribution as described by the array. One thing to be emphasized is that the data element is -1 or any negative value if the agent does not know existence of the exit. In Table 6 the agent named by Luca does not know there is Exit2 in the given compartment layout. Thus, the in the door selection algorithm the probability of Luca's using Exit2 does not dynamically evolves until he is told by others that there is Exit2 or he actively find this exit.*

***Table 8. Data Array of Exit-Selection Probability***

| *&Ped2Exit* | *Exit0* | *Exit1* | *Exit2* |
|---|---|---|---|
| *agent0* | *0.7* | *0.2* | *0.1* |
| *Luca* | *0.3* | *0.5* | *-1* |
| *Irving* | *0.0* | *-1* | *0.5* |

*In addition all the exits are classified in two types for each agent: known exits and unknown exits. An unknown exit will not be included in the exit-selection algorithm for the specific agent until the agent gets a way to know it. When the element of the data array is assigned with zero probability, it means that the agent does know the exit, but never goes there before. This case usually fits in any special passageway for fire egress or emergency use. For example, in Table 6 Irving does know Exit0, but he never use it previously, and thus the historical usage frequency of that exit for Irving is simply zero, i.e, assigned with zero probability initially. This exit is definitely included in the computational loop of exit-selection algorithm, and thus the probability for Irving to select Exit0 is dynamically updated through the simulation process. This algorithm is being tested and it is inspired and developed based on the codework of FDS+Evac.*

*In the simulation process the prior probability distribution of using multiple exits will be dynamically updated when new information received. We have briefly talked about this issue when introducing the parameter <p>, <pMode> and <p2>, where a more realistic and complicated scenario is considered by using solver=2 and it assumes that each individual's choice is made by integrating three major factors (See Figure 4): (1) historical knowledge of*

*building layout or prior information such as habit of using different path; (2) Judgment on current situation in egress (e.g., the distance to an exit) (3) Choices of other individuals. Because each individual's decision is interacting with others, the collective decision-making is studied with integration of various social-psychological findings in evacuation research [Ozel, 2001; Staal, 2004; Santos and Aguirre 2005]. As a result, the data block claimed by &agent2exit only gives the initial value of exit-selection probability distribution, and the input parameters <p> <pMode> and <p2> critically determines how it evolves dynamically in the simulation. To fully describe this scenario users should well define parameter <p> <pMode> and <p2> for each agent in csv data file. As shown in Figure 4, each agent integrates new information with his/her past experience, and there are two source of new information: (1) Information from egress facilities and corresponding weight is given by <p2>; (2) Information from surrounding people and corresponding weight is given by <p>. Consequently the probability of an agent exit-selection is summarized as below.*

*Prob(individual i select an exit) = (1-p)(1-p2)*Prob(individual i select an exit by using prior experience)*
*+ (1-p)*p2*Prob(individual i select an exit by utility of the exit)*
*+ p*Prob(individual i select an exit by hearing others' opinions)*

$$0<=p<=1,\ 0<=p2<=1$$

*After an exit is selected by an agent, the evacuation route will be generated towards the exit selected. There are a number of methods used in existing egress simulators. Our algorithm is partly learned from FDS+Evac, where the route is calculated by a two-dimensional flow solver. The computation result is a 2D flow field that guides agents to the exit selected. The flow field can be better explained as a social field related to social norms or other behavioral characteristics (Lewin, 1951; Helbing et. al., 2000; Wang, 2022), and we will further elaborate the idea in future. In this algorithm each exit is a sink point, and solver calculates the route as the crowd flow move to the sink (Korhonen et. al., 2008; Korhonen, 2016). The detailed discussion of the flow solver will not be included in this article, but we emphasize that this method is more suitable to describe human behavior on the background of social science and psychological theory.*

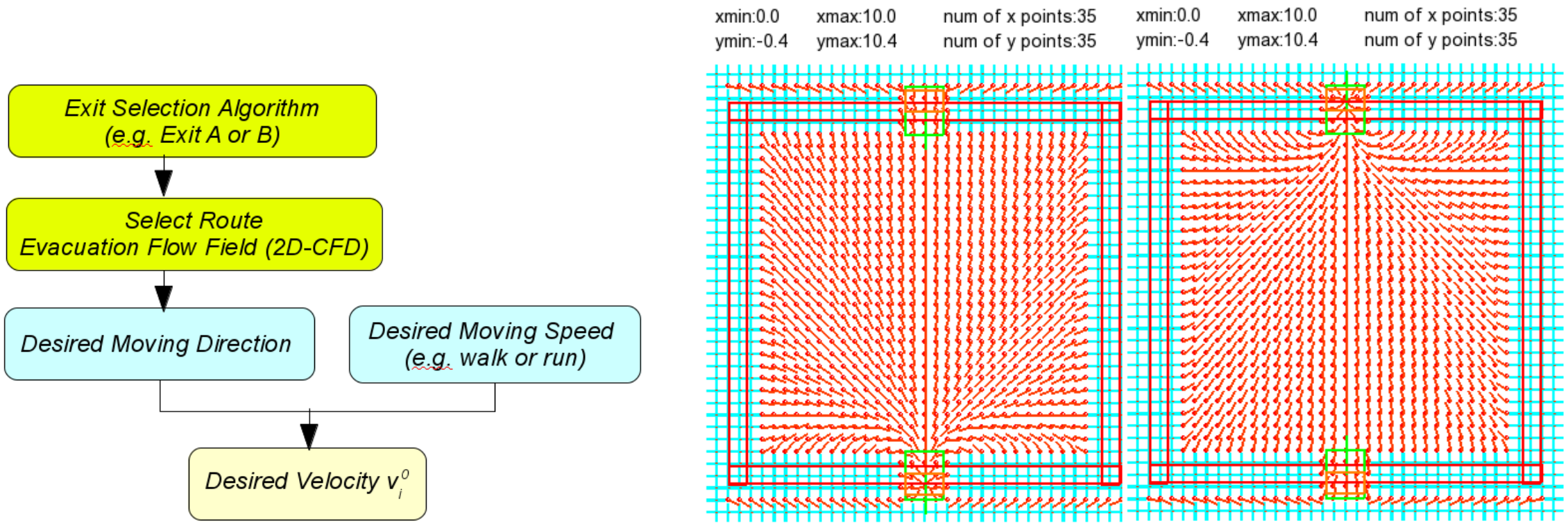

***Figure 8. Simulation of crowd egress with exit-selection algorithm.***

*In an addition, there is another data array called exit2door, which is claimed by &exit2door in the input csv file. This data array specifies the door direction towards a given exit. For example, in Figure 8 there are two exits in the structural layout, and if the upper right one is selected by an agent, the door direction towards this exit will be given by the first row in the following data array. If the lower left exit is selected, the door direction will be given by the second row. The data array &exit2door is optionally specified by users. In fact if a flow solver is used in our program, this array could be automatically generated by the egress flow field. Thus, users could usually omit this data array in the input csv file unless no egress flow field is calculated (solver=0) in the program.*

***Table 9. Data Array of Exit2Door Matrix***

| &Exit2Door | door0 | door1 | door2 | door3 | door4 | door5 |
|---|---|---|---|---|---|---|
| Exit 0 | 1 | 1 | -1 | -1 | -2 | -2 |
| Exit 1 | -1 | -1 | -1 | -1 | 2 | 2 |

## *4. Configure Simulation Parameters and Run Simulation*

*As above we briefly introduce how to write a csv file to describe agents, walls, doors and exits. In addition another config.txt file is often used jointly with csv file to configure the simulation parameters. In our program there are mainly two methods to configure a simulation object and launch the program.*

*(1) The first alternative is using GUI (Graphic User Interface), and it is an easy method to setup all the simulation parameters. The GUI is written in ui.py as an independent component in our program, and it is developed by Tkinter, which is being tested in the past two years. There are also several other versions of GUI in trial, and users' feedback are important for us to improve the GUI development. When the GUI is started, users can configure the simulation object by selecting items in each panels, and then start the simulation process.*

*(2) The other alternative is using a script file config.txt to setup the simulation parameters, and it is to be placed in the same folder as the input csv file. The configuration file config.txt is also useful to start the program without using GUI, i.e, starting the program through main.py, and this method is useful to start the program in a small or tiny Linux system (e.g., raspberry pi OS, Arch Linux). We will briefly introduce this issue in Section 5.2. The reading process of config.txt is packed into a function called readconfig() in simulation.py. User are welcome to check or modify readconfig() for their preference or convenience.*

### ***4.1 In Tkinter GUI:***

*When tkinter window (GUI) is activated, users will see several panels including <QuickStart>, <Parameters> and so forth. Next, we will briefly introduce how to set up the simulation in the panels.*

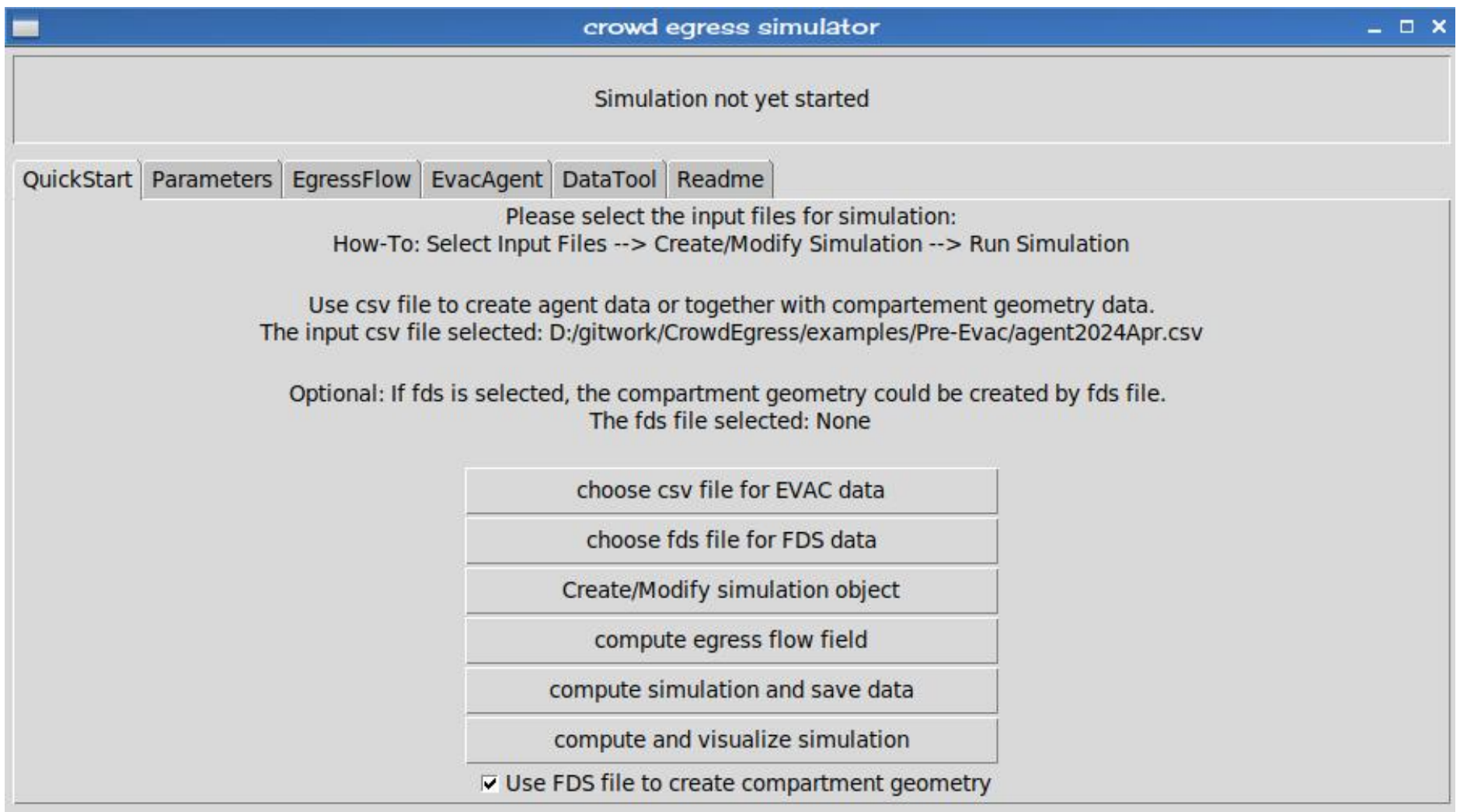

***Figure 9. User Interface of CrowdEgress (QuickStart)***

*The default first panel is called <QuickStart>. In this panel users select the input files to get a quick run of the simulation. Choose csv file to specify agent data. Users can optionally use fds file to create the compartment geometry, and the pedestrian features are described in csv file. If both csv and fds files are presented, the compartment structure will be created by fds file. If fds file is omitted, the compartment geometry should be described in csv file. The agents must be specified in csv file currently while the walls, doors and exits can either described by csv file or fds file. Please take a look at the examples for details.*

*<Create/Modify simulation object>: After the input files are well selected, it is suggested that users first create the simulation object to see if the structural layout is created as expected. Another screen is next displayed by pygame, where users could check all the entities in the simulation, such as wall, doors, exits and initial potions of each agent. We call this screen TestGeom in this manual, and it means testing the geometric settings of compartment layout. In TestGeom user could add walls, doors or exits, or change the direction of doors or exits, or change the initial position of each agent. Whenever users change the input files or after an existing simulation is complete, users should re-click this button to initialize the simulation object. Otherwise the simulation object will remain to be the previous one in the memory and new simulation object will not be created.*

## 4.2 In Pygame Screen (TestGeom):

*When pygame screen is activated, press keys of <PageUp/PageDown> in your keyboard to zoom in or zoom out the entities in screen. Use direction keys to move the entities vertically or horizontally in screen. In fact there are several sections in pygame screen, and these keys are functional in similar ways in all the sections. The first section to be introduced is called TestGeom as shown in Figure 5 and Figure 10, where users can visualize compartment geometric layout and modify it manually.*

*First of all, users can also visualize and check the mesh field in TestGeom. Please select the item <ShowData> -> <ShowMesh> in the menu bar or simply press the number <5> and the mesh field will be displayed in the pygame Screen of TestGeom. Press the number <5> again and the mesh field will be hidden. When the mesh is displayed, the detailed information is also listed on the right side of the screen, and users can check if the mesh is created as expected or not.*

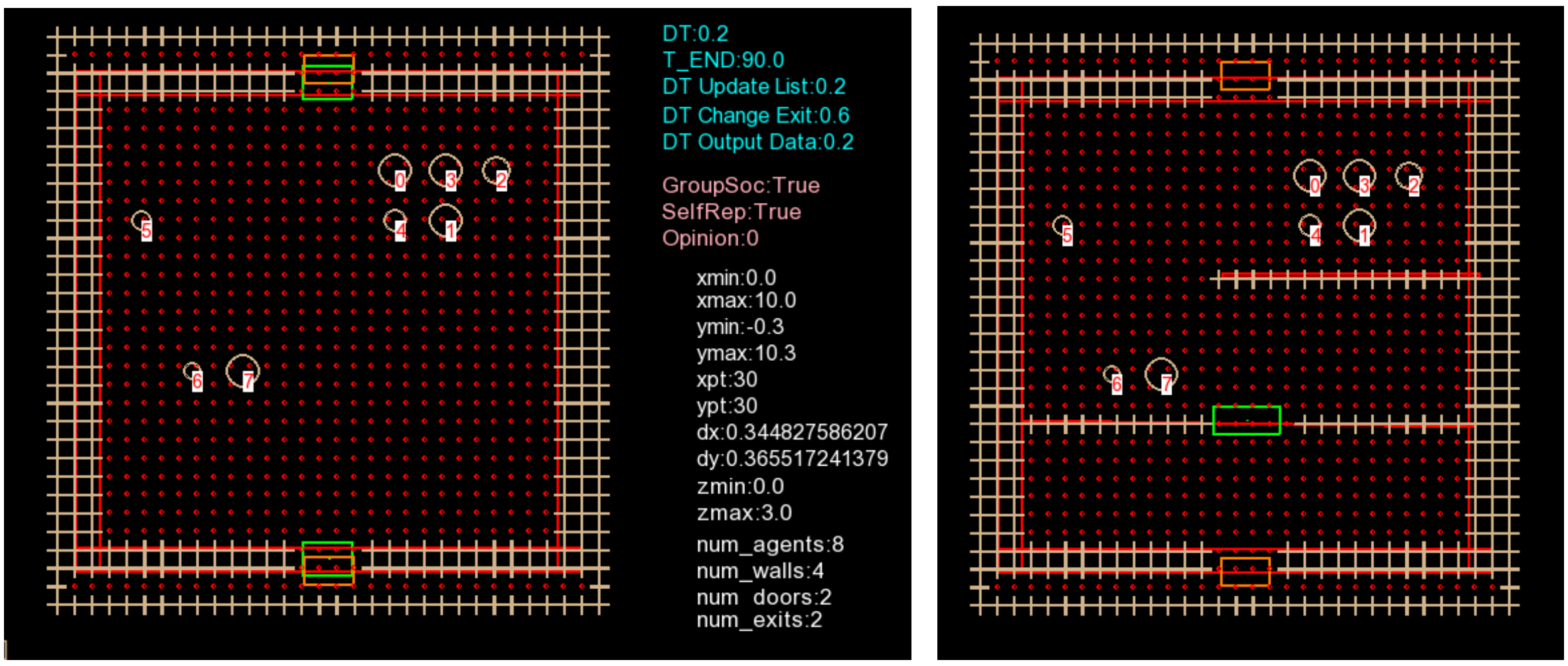

*(a) Showing Egress Mesh in TestGeom* *(b) Adding two walls (line shape) and a door*

***Figure 10. Visualization of Mesh Field in Pygame Screen of TestGeom.***

*In TestGeom users can add walls, doors or exits by selecting the corresponding items in the menu bar (See Figure 5). When the corresponding item is selected, simply drag a rectangular region to create a new wall, door or exit in rectangular shape. A wall in line shape can also be created by drawing a line from its start point to end point. If users are not satisfied with the new entities just created, please press <z> in the keyboard to remove it. Please be careful to remove entities because the last element in the list of walls, doors and exits will be removed by pressing <z>, not only for the new entities that users graphically created in testGeom. In fact, all of the walls, doors exits are sequentially stored in lists, and pressing <z> will delete the last element in the list (list.pop). Thus, all the element of walls/doors/exits can be removed if users keep pressing <z> in TestGeom until the lists become empty. Adding any walls/doors/exits will simply add elements in the corresponding lists. Sometimes it is necessary to adjust the exit-selection probability if new exits are added in testGeom. Namely, the number of column in data array &Agent2Exit (See Table 1) should be equal to the total number of exits. If users do not manually update &Agent2Exit in the input csv file, the program will automatically add zero columns there.*

*When all the items in the menu bar are closed, users may also adjust the initial position of each agent or change the direction of a door or exit. Simply drag an agent to a new position, namely, drawing a line from the existing position to a new position, and the agent will be moved there. The direction of a door and exit is adjusted by drawing a line from outside to inside of the door or exit. However, if solver 1 or 2 is selected for the simulation, users do not need to care about this issue because such directions will be automatically updated by the egress flow solver.*

*Users can also dump geometric data (e.g., wall data and door data) into csv file by selecting the item <OutputData> in the menu bar in TestGeom. The output file is saved as bldDataRev.csv for any modification of the compartment geometric layout in TestGeom. The data can also be briefly shown in the screen by selecting the item <ShowData> in the menu bar. After the simulation object is initialized or modified, users can click <Simulation> and the simulation starts and the program goes to the second section called RunSimulation. There are several options for users to configure how the simulation is supposed to run.*

*<Compute Egress Flow Field>: The first option is to compute the egress flow field and display it on the pygame screen, and agent-based simulation is not launched in this option. In the pygame screen users will find how the mesh is created such as the number of x points and y points, the boundary value of mesh (i.e., min_x, max_x, min_y, max_y). Users will also see the flow field towards each exits. If solver 1 is selected, there is only one flow field computed, which is simply using the nearest-exit strategy (See Figure 7). If solver 2 is selected, there are several flow field computed, which corresponds to the roadmap towards each exit (See Figure 8). The nearest-exit strategy is also additionally computed in solver 2. Users could press <o> and <p> in the keyboard to switch from different flow field. Please also see the third panel of EgressFlow for more details.*

*<Compute Simulation and Save Data>: The second option is to compute the agent-based simulation with the flow field. The numerical result will be written into a binary file (.bin), a text file (.txt) and a npz data file (.npz). but not displayed online in pygame screen. If your input csv file is named by agent.csv and simulation is complete at time of 2024-03-02_23_09_27 (in format of year, month, date, hour, minute and second), the output data files will be generated as agent_2024-03-02_23_09_27.bin, agent_2024-03-02_23_09_27.txt and agent_2024-03-02_23_09_27.npz. We will introduce these files in detail in the following section.*

*<Compute and Visualize Simulation>: The third option is to compute the agent-based simulation and visualize the numerical result online in pygame screen. This is the most common and important option such that users are able to directly observe how agents interact and move towards a selected exit in the compartment layout. As below we will introduce this pygame section as called RunSimulation. Users can directly observe the forces and movement trace of each agents, pause the simulation. The output data can also be saved as in the previous option, namely the numerical results are saved into a binary file (.bin), a text file (.txt) and a npz data file (.npz).*

### ***4.3 In Pygame Screen (RunSimulation):***

*In pygame section of RunSimulation the agent-based simulation is visualized on the pygame screen. User can pause the simulation, but cannot rewind it in current version (Version 2.3). A binary data file is optionally created when the simulation runs and users can also visualize the data after the simulation is finished. In both phases of TestGeom and RunSimulation, there are hot keys defined. Use pageup/pagedown to zoom in or zoom out the entities in screen. Use space key to pause the simulation. Use direction keys to move the entities vertically or horizontally in screen. Use 1/2/3 to display the door or exit data on the screen. Press I to show agent index number. Press S to show their stress level timely. The stress level is computed based on our work Wang and Wang 2020.*

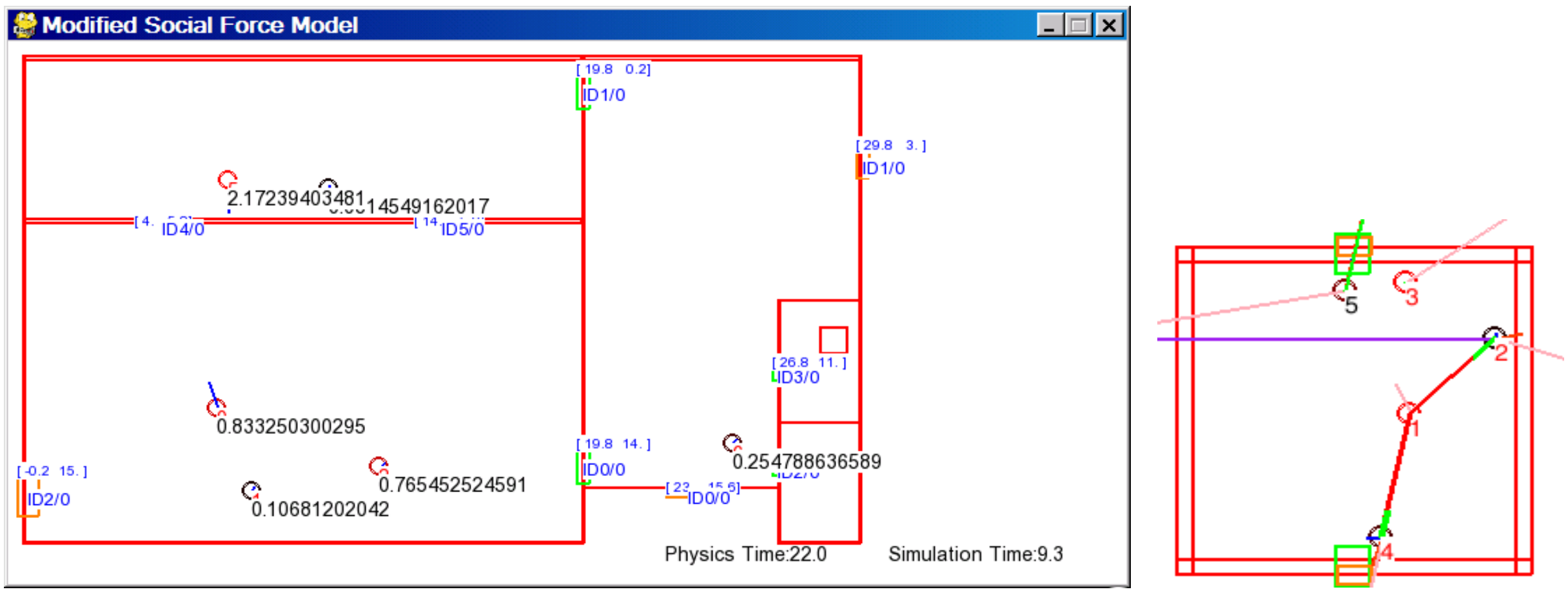

***Figure 11. Agents with forces or stress level indicated.***

*The second panel is called Parameters, where users specify the basic parameters before the simulation is launched. An important issue is that z-interval (min_z and max_z) should be specified when a FDS+Evac input is used to create the compartment geometry setting. The reason is that FDS+Evac is a 3D simulator, and its input file may include several computation meshes as several floors in a building. However, our simulator CrowdEgress is a 2D simulator only for single-floor layout. Thus, if a FDS+Evac input file is used with several compartment floors, sometimes users should check or modify the min_z and max_z to determine which floor should be computed in the simulation. The*

default value in CrowdEgress is given as min_z=0.0 and max_z=3.0, and it is generally suitable for most FDS+Evac file with only one compartment floor. Other parameters are briefly introduced as below.

<Use nearest exit strategy>: Solver 1 is used if selected. Namely, each agent is guided to the nearest exit and exit-selection algorithm is not involved. Solver 2 is used by default.

<Show Stress Level in Simulation>: Indicate the difference between actual velocity and desired velocity. Please refers to Wang, 2021 for more details.

<Show compartment data in simulation>: Show the geometric data of doors and exits in the pygame screen such that users can identify each door and exit by their indices and default directions.

<Show forces on agents in simulation>: Visualize forces in the pygame screen. The forces are measured by several lines shown on each agent. The line in purple color is the wall force; line in green color represents the door force. The line in pink color is the social force.

<Write data to a binary/npz file>: Write simulation data into a binary file which is compatible to FDS prt5 data format. Also, write the array-like or matrix-based data into a npz data file, which is a special form used in Numpy. These data files are used to visualize the numerical result after simulation is accomplished.

<dtSim>: The time interval (second) for each simulation step. In Figure 12, for example, the time interval is 0.2 second for each simulation step, and all the agents state including position, velocity and forces are updated numerically every 0.2 second.

<dtDump>: The time interval to write simulation data into the binary and npz data file. For example, in Figure 12 the simulation data is written into the binary and npz file every 0.2 second, which is the same as the default time interval for each simulation step. This means that all the simulation data in all time steps are saved into the binary and npz data files.

<tEnd>: The entire time span for simulation. In Figure 12, for example, the simulation starts at t=0.0 second and is accomplished at t=19.0. This value is only valid if simulation is computed without timely visualization, namely, if users selects the mode of <Compute Simulation and Save Data> as introduced before.

<Use config.txt to overwrite parameters selected in GUI panels>: If checked, all the above simulation parameters will be overwritten by reading the script of config.txt, which is placed in the same folder of the input csv file.

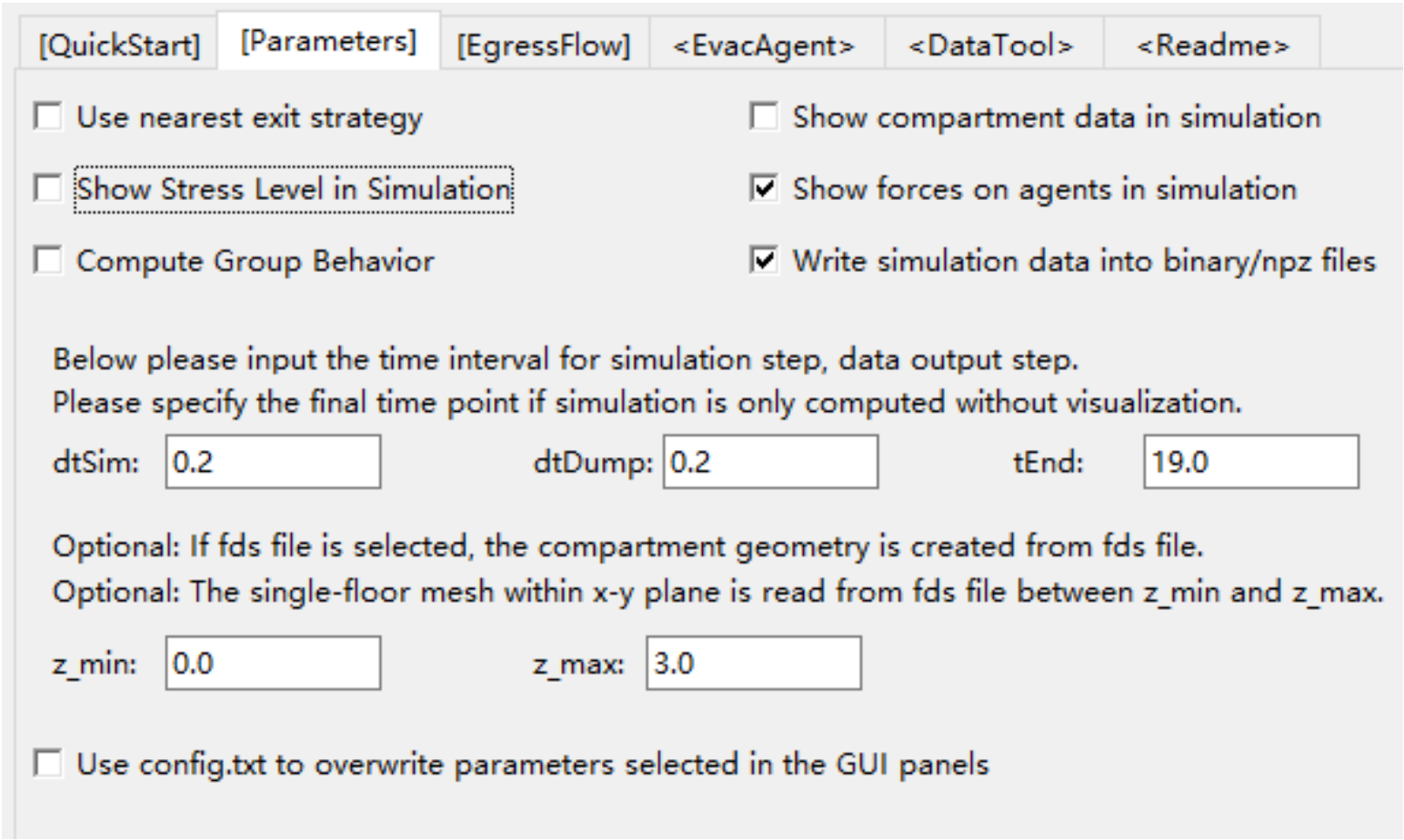

***Figure 12. User Interface of CrowdEgress (Parameters)***

The third panel is called EgressFlow, where users specify the basic parameters to calculate the egress flow field. The calculation is executed based on the solver selected. If solver=1, the egress flow field is calculated only for the nearest-exit strategy. If solver=2, several flow fields are calculated, and each one corresponds to an exit, and the total number is the number of exits plus one because the nearest-exit strategy is also calculated for solver=2.

In the panel of EgressFlow users are expected to input the dimensional measure of the calculation mesh, including the minimal and maximal value in x axis (x_min and x_max) and minimal and maximal value in y axis (y_min and y_max). The number of points are denoted by num_x and num_y. When the number of points increases, the computational mesh are refined and the calculation time is supposed to be increased also.

*There are two buttons in this panel. One is <Compute Egress Flow Field>, and it is the same as in panel of QuickStart, and it generates a pygame screen, where users will find how the mesh is created such as the number of x points and y points, the boundary value of mesh (i.e., min_x, max_x, min_y, max_y). Users will check the flow fields in the pygame screen: if solver 1 is selected, only one flow mesh is computed for the nearest-exit strategy (See Figure 7). If solver 2 is selected, several flow fields are computed, which corresponds to the roadmap towards each exit (See Figure 8). The nearest-exit strategy is also additionally computed in solver 2. Users could press <o> and <p> in the keyboard to switch from different flow field.*

*Another button is especially useful for Solver 1 and it shows the computational result of crowd fluid dynamics. This fluid-based model is different from the agent-based model, and it assumes crowd movement as mass flowing in a two-dimensional space, and it is basically an analytical model at the macroscopic level, aggregating many particle-like individuals into fluid-like model of crowd. In a sense the fluid-based model (macro-level) is derived from homogeneous individual equation (micro-level) based on physics laws and mathematical principles. The resulting model is a set of partial differential equations (PDE). The analytical solutions to these equations are often difficult in mathematics, but numerical solutions are obtained nowadays by advanced computing methods. Thus, we apply a simple version of fluid-based model in our program by extending 1D traffic problem (Lighthill and Whitham, 1955) to 2D crowd fluid problem. The computation process of this simple fluid-based model is included in Solver 1. In other words when users select <Use nearest exit strategy> in the <Parameters> panel, it means that Solver 1 is selected and this fluid-based model will be calculated along with the egress flow field.*

*The solution could be visualized by using the second button in the panel of <EgressFlow>, i.e.,the button named by <Read output npz file and show crowd fluid model>. The computational result is stored in vel_flow1.npz in the example folder, and it is visualized in Figure 14 and Figure 15.*

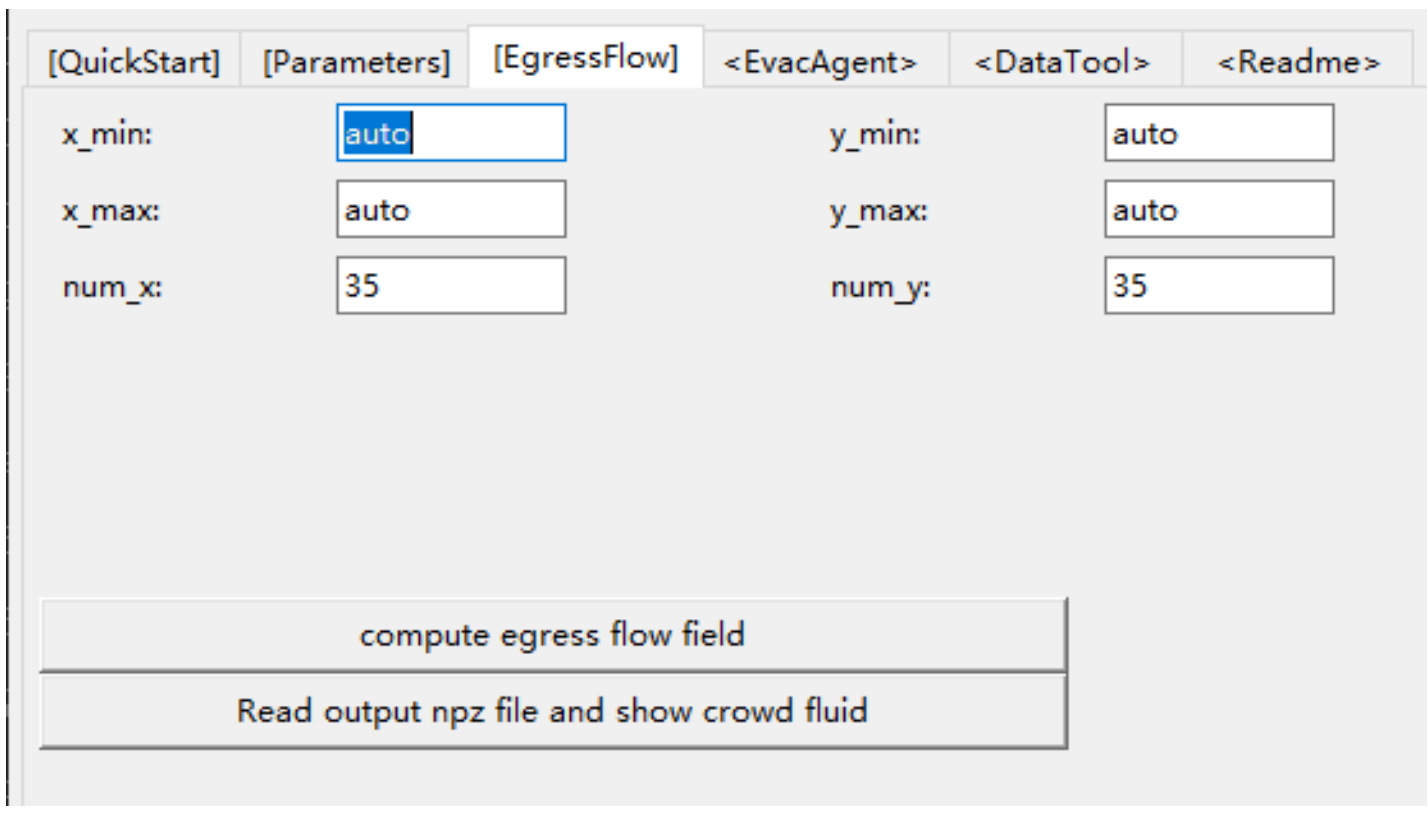

***Figure 13. User Interface of CrowdEgress (EgressFlow)***

*As shown in Figure 13 the basic information of egress flow field is first displayed such that users could check if the mesh is generated as expected, including the number of x points and y points and the boundary value of mesh (i.e., min_x, max_x, min_y, max_y). When the pygame screen is next displayed, the crowd flow density is denoted by a small black dash in the mesh, and users will see how they move towards the nearest exit. In this process it is also easy to use mouse to check the numerical value of the flow velocity and density for each mesh unit. U and V are the flow velocity, and R represents the flow density. Ud and Vd are the desired flow velocity, which always point towards the nearest exit. BLD means whether the mesh unit is free for crowd flowing or solid boundary generated from the compartment layout.*

*In this process users could also use <pageup/pagedown> to zoom in or zoom out the entities in screen. Use <space> key to pause the simulation. Use arrows to move the entities vertically or horizontally in screen. The general framework of the fluid-based analysis is illustrated as below.*

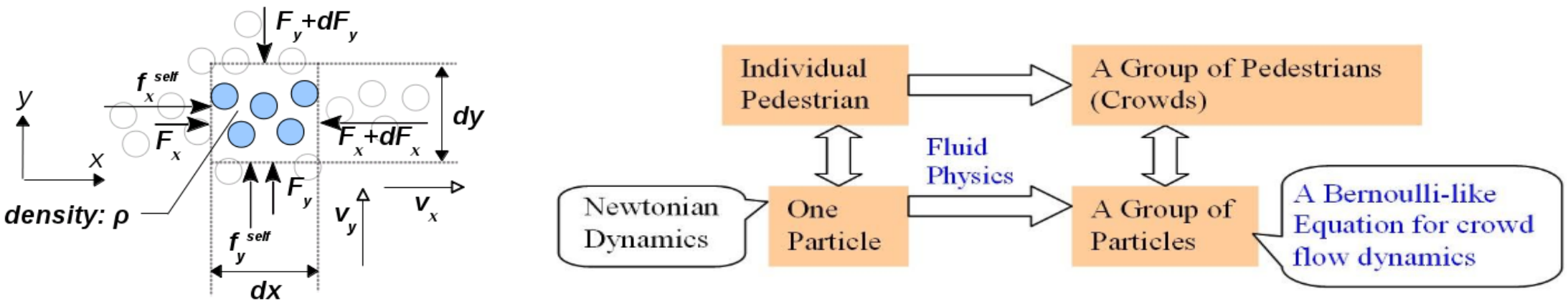

***Figure 14. Fluid-Based Analysis of Crowd Movement***

*The CrowdEgress program implements a macroscopic crowd dynamics and evacuation simulation based on fluid dynamics principles. The flow solver is packed in flow.py and simulation.py in the source code. The code bridges computational fluid dynamics (CFD) and pedestrian dynamics. It operates in two main phases: Static Navigation Field Generation and Dynamic Crowd Fluid Simulation.*

***Phase 1: Static Navigation Field (Poisson Equation)***

- ***Geometry Rasterization (****build_compartment****)****: The physical layout (walls, doors, exits) is mapped onto a 2D grid (meshBLD). Walkable areas are 1.0, and walls/obstacles are 0.0.*
- ***Sink Generation (****build_sink****)****: Exits are treated as "sinks" (assigned a large negative value, -200.0) in the right-hand side (b) of Poisson's equation.*
- ***Potential Field Solver (****possion_func****)****: It solves the 2D Poisson equation* $\nabla^2 p=b$ *using the* ***Jacobi iterative method****. Walls are enforced by multiplying the pressure p by the BLD matrix (setting p=0 at walls).*
- ***Gradient Calculation****: The desired velocity field (Ud,Vd) is computed as the negative gradient of the potential field* $(-\nabla p)$*. This creates a vector field that naturally routes pedestrians around obstacles toward the exits.*

***Phase 2: Dynamic Crowd Fluid Simulation (lwr2D)***

*This function solves the 2D conservation of mass equation (Lighthill-Whitham-Richards or LWR model):*

$$\partial\rho/\partial t+\nabla\cdot(\rho v)=0$$

- ***Density (ρ /*** *R****)****: Represents the crowd density.*
- ***Velocity (v /*** *U, V****)****: The speed is determined by a fundamental diagram:* $v=v_f(1-\rho/\rho_m)$*, where* $v_f$ *is free-flow speed and* $\rho_m$ *is maximum density. The direction is given by the normalized static navigation field (Ud,Vd).*
- ***Numerical Schemes****: The code includes multiple modes for solving the PDE:*
  - *mode=0, 1: Lax-Friedrichs scheme (a classical finite-difference method for hyperbolic PDEs).*
  - *mode=2: Standard central-difference scheme with driving force terms.*
  - *mode=3, 6: Upwind/Donor-cell scheme for the LWR model, ensuring numerical stability for advection-dominated flows. mode=3 specifically uses an exit_pt matrix to drain density at the exits.*

***Phase 3: Visualization (draw_vel)***

*Uses pygame to render the vector field and the building layout interactively. Users can zoom and pan using keyboard inputs.*

*Theoretically, our program treats a crowd of people as a continuous fluid and calculates their movement through a 2D compartment (e.g., a room with walls, doors, and emergency exits).Last but not least, analog of fluid dynamics is only valid to crowd at medium or high density, where continuity hypothesis holds for crowd flowing in a planar space. If there are only sparse individuals, continuity hypothesis cannot hold, and fluid-based analysis is not actually suitable for such low-density crowd. The resulting fluid model provides a practical perspective to explain crowd behavior at bottlenecks (e.g., narrow passages), where crowd density are sufficiently large and short-range physical interactions are dominant among people. Such interactions are among the major cause of crowd disastrous events like stampede.*

*In the past 20 years of evacuation study, there is also a branch to use cellular automata (CA) to simulate evacuees motion in 2-dimensional lattice. Either deterministic or probabilistic rules can be applied. CA model can also be applied with discrete evacuation mesh as developed above. If anyone is interested to test CA model in our simulation platform, please contact me and I am glad to provide you guidance and help.*

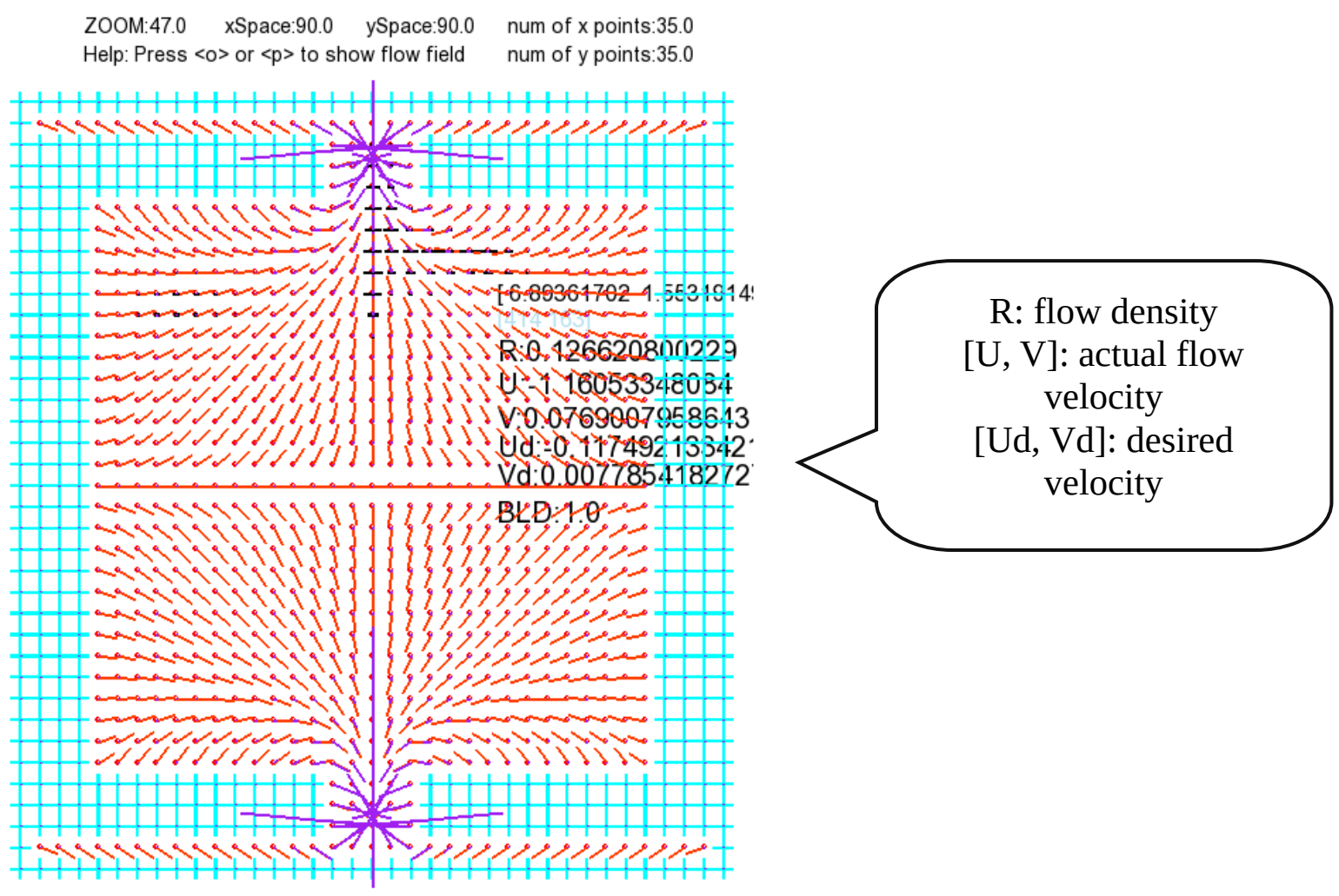

***Figure 15. Computational Result of Crowd Fluid Model***

*The fourth panel is called EvacAgent, where users specify the basic parameters to calculate the agent-based simulation. Many parameters in this panel refers to the theoritical model of agents, and if users are interested, please browse our manuscript Wang et. al., 2004 for more details of the theoritical model.*

*<Selection of Short-Range Force>: Users could choose the short-range forces for agent interaction. Such forces are effective only if agents are sufficiently close to each other. As shown in Figure 16, value 0 is for social force in Helbing, I. Farkas, T. Vicsek 2000 and value 1 is for magnetic force in Okazaki, 1979.*

*<dtAtt>: The time interval (second) to update the attention list for all the agents. In Figure 16, for example, the time interval of dtAtt is 1.0 second, and the attention lists for all the agents are upated numerically every 1.0 second.*

*<dtExit>: The time interval (second) to upate the target exit for all the agents based on the exit-selection probability. In Figure 16, for example, the time interval of dtExit is 1.0 second, and all the agents will reselect the target exit numerically every 1.0 second. This parameter is not valid for solver=1, where the nearest-exit strategy is simply used.*

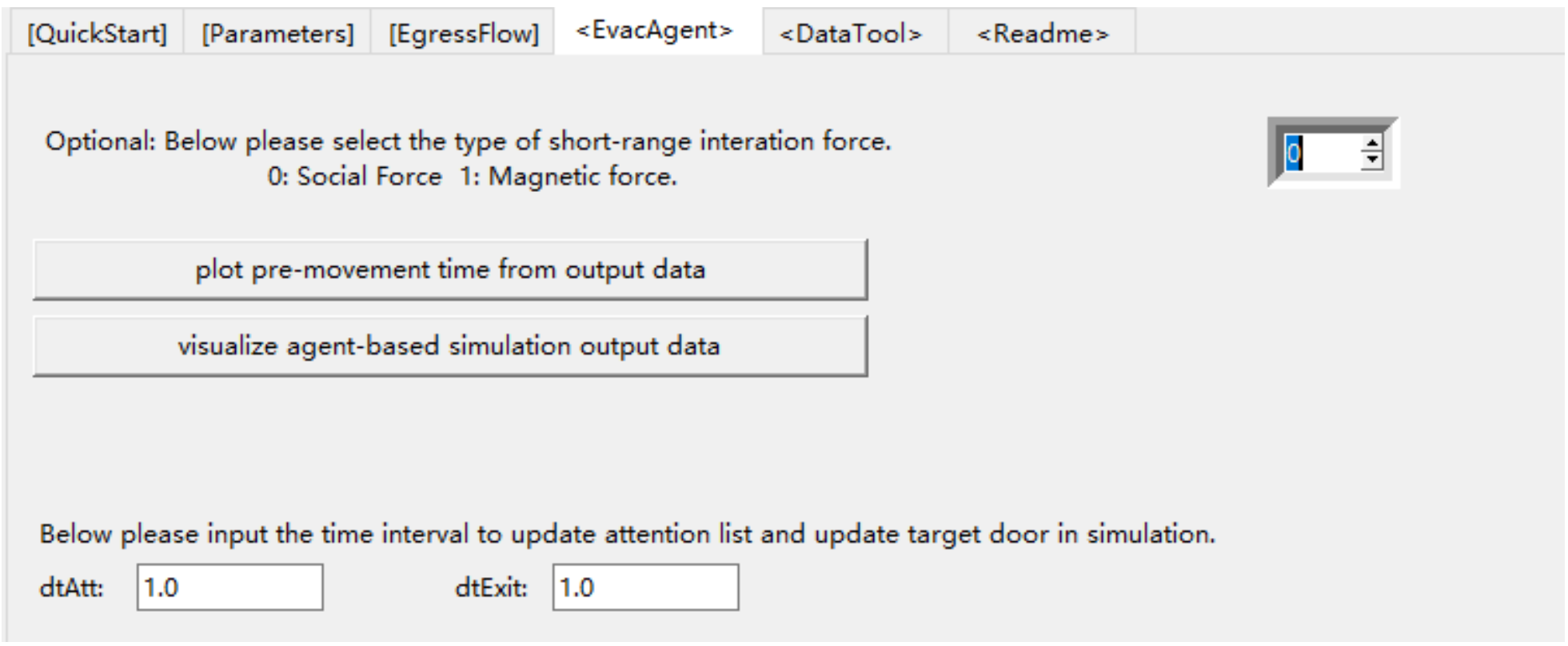

***Figure 16. User Interface of CrowdEgress (EvacAgent)***

## *5. Analysis and Visualization of Output Data*

*The output data of above agent-based simulation are classified in two forms, namely, the non-matrix data and matrix-based data. The non-matrix data is used to store agents' position velocity and forces for all the time steps. The social relationship and egress flow field are matrix-like data in our program. Both forms are also written into a text file in the simulation process. Users could easily check the text data by using a text editors. The output data files are listed as below.*

***Table 7. Output Data of Simulation.***

| | |
|---|---|
| *Text Data* | *The text file is a log of computation process and it records agents' position velocity and forces for all the time steps. It also includes other agent features such as exit selection probability, dynamics of pre-movement time, the matrices of social relationship, and so forth. In a sentse the text file contains all the data, which are included in the binary file and NPZ file.* |
| *Binary Data* | *The binary data file mainly records agents' position velocity and forces for all the time steps. It also record other non-matrix data for each agents, such as pre-movement time, stress level, exit selected and so forth. The binary data could be visualized in pygame after the simulation is complete. Use a button in the panel of EvacAgent and select the binary file (suffix of .bin file), and the data will be extracted and displayed in pygame section.* |
| *NPZ Data* | *The NPZ data file is a special form in NUMPY, and it is useful to store matrix-like data. In our program the NPZ data file is used to store the social relationship of agents as they move and interact. For example, the see list, attention list and talk list are all saved in the output npz file. In addition, the computational results of egress flow fields are also saved in NPZ form. The filename are vel_flow1.npz and vel_flow2.npz for Solver 1 and Solver 2, respectively.* |

*The output data files for egress flow field are vel_flow1.npz and vel_flow2.npz, for Solver 1 and Solver 2, respectively. Such data files are stored in the example folders and they are not recalculated unless the compartment layout are changed. The agent-based simulation data include a binary data file, a npz data file and a text file, which all have the same prefix name. If your input csv file is agent.csv and simulation is complete at time of 2024-03-02_23_09_27 (in format of year, month, date, hour, minute and second), the output data files will be generated as agent_2024-03-02_23_09_27.bin, agent_2024-03-02_23_09_27.txt and agent_2024-03-02_23_09_27.npz.*

*Users could direct open the text file by using any text editors and check the agent data there, including the agent's position and velocity and forces at each time step. Such data are 2D vectors, and in the text file we also list the vector's norm right before the vector. The text output file is illustrated as below.*

```
&SimulationTime:0.0

Agent:  0:0
Simulation Time:0.0
Position:       [ 6.38051627  2.13204704]
Velocity:       0.331895503341: [-0.319925882733 , 0.088329240321]
DesiredVelocity:        1.40009118553: [1.38003892471 , -0.236109919532]
@motiveForce:   75.167947369: [58.8391268948 , -46.777959104]
@socialForce:   280.880288272: [-193.147180538 , -203.931123152]
@wallForce:     0.0: [0.0 , 0.0]
@doorForce:     50.2233085123: [-0 , 50.2233085123]
@diss:  0.329424300315: [0.317543802224 , -0.087671564987]
@selfRepulsion: 47.5152048969: [-37.1934217737 , 29.5693096497]
@subF:  242.766733406: [-171.752176136 , -171.571783348]
@objF:  0.0: [0.0 , 0.0]
Premovement Time:2.4
ExitSelected:1
numOtherSee:4
numOther:1

Agent:  1:1
Simulation Time:0.0
Position:       [ 7.61065949  3.20015733]
Velocity:       1.88782302369: [1.46416279368 , 1.1916807804]
DesiredVelocity:        1.43993164818: [1.42851441541 , -0.180968827126]
@motiveForce:   92.8415883224: [1.98065540938 , -92.820458555]
@socialForce:   94.1034870795: [18.6760508234 , 92.2316182563]
@wallForce:     0.0: [0.0 , 0.0]
```

***Figure 17. Output Data in Text File.***

*As for the npz data file, it is a special form used by python numpy package, and it is easy to store the data array and matrix. This matrix-based data could be visualized together with the binary data file in our program.*

*<plot pre-movement time from output data>: Select the output binary file and the pre-movement time for each agent is plotted as exemplified in Figure 18. In Figure 18 each colorful line represents an individual's $tpre_i$ and the value of $tpre_i$ dynamically changes as agents interact with each other. When the colorful line drops to t axis, it means that the individual reaches an exit and thus is removed from computational loop. In particular there is a*

gray slash line which has angle of π/4 to the t axis, and this gray line divides the t-tpre plane equally into two regions. The upper left region indicates the pre-evacuation phase because $tpre_i$ is larger than simulation time ($tpre_i$>t). The lower right region represents the movement phase since the simulation time t reaches $tpre_i$ and thus individuals start to move to exits.*

*In particular it is noted that individual agent 4 (purple line) sways the opinion between individual agent 1/2/3 and individual 7. Agent 4 initially communicates with agent 1 and agent 2, implying that agent 1 and agent 2 are probably in his or her attention list, or even in the talk list. Afterwards, agent 4 communicates with agent 7 for quite a while, and after 7 seconds agent 4 comes back to communicate with agent 1/2/3, and his or her pre-movement time converges around 6 seconds. Here when the colorful line drops down to the x axis, it means that the corresponding agent reaches an exit and thus becomes out of the computation loop. In the following figure, it shows that agent 5 first gets out at 10 seconds, and then agent 0 next moves out at around 11 seconds, and agent 3, agent 2 and agent 1 reaches an exit sequentially at around 12.5 seconds, 13.5 seconds and 14 seconds.*

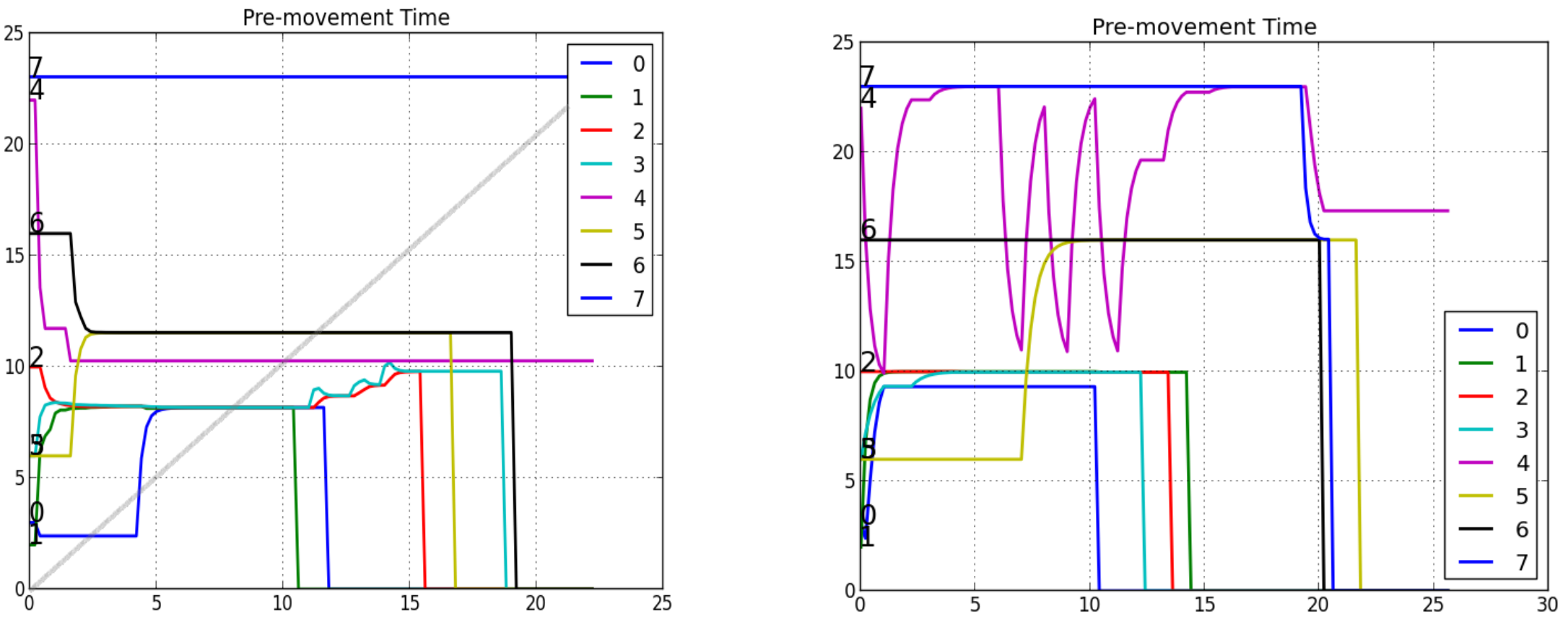

***Figure 18. Visualization of pre-movement time from output data***

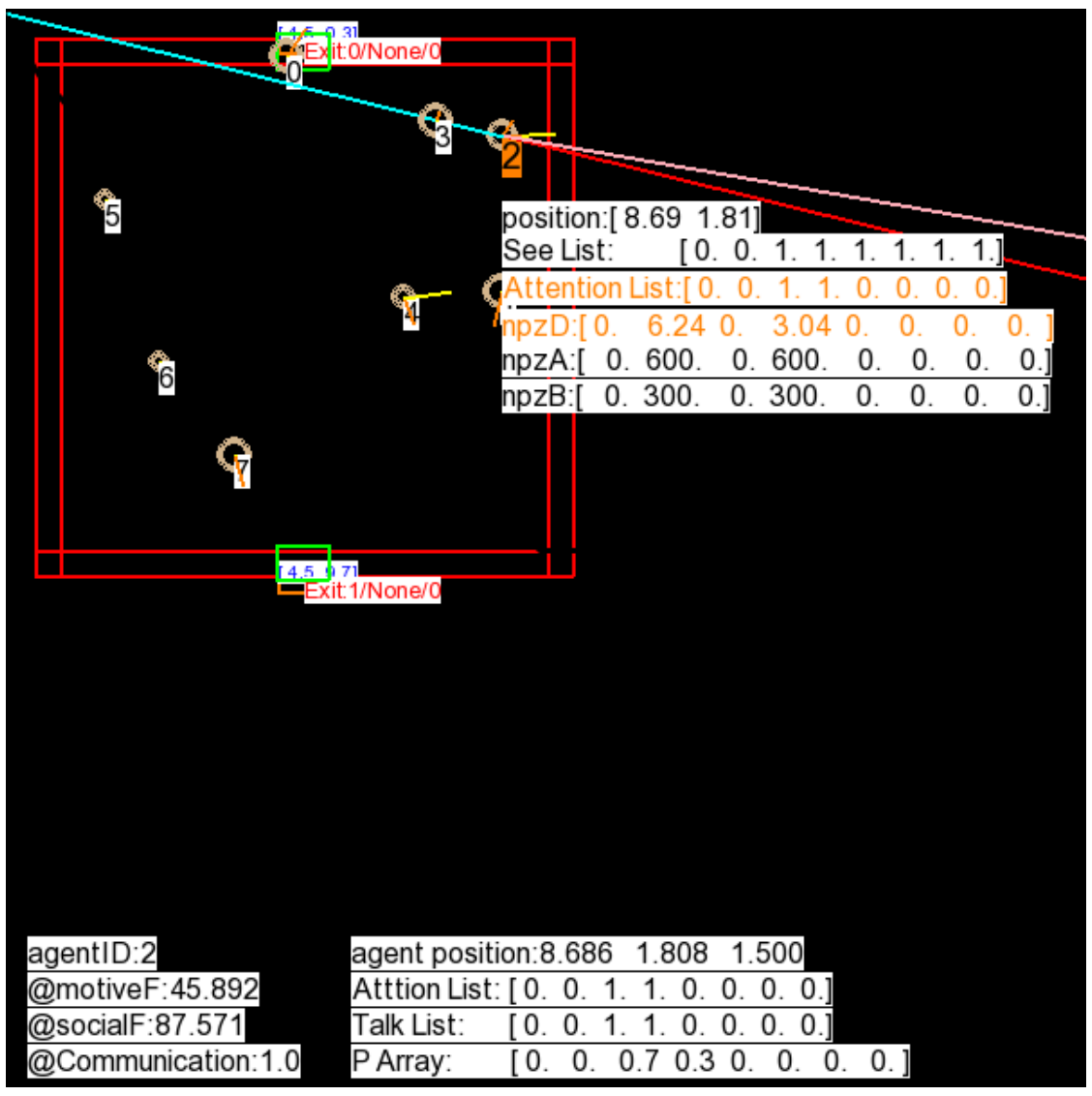

***Figure 19. Visualization of agent movement from output data***

*<Visualize agent-based simulation output data>: Select the output binary file and the data will be extracted to show the simulation by steps in Figure 19. Similar to testGeom and runSimulation as introduced before, press <pageup/pagedown> to zoom in or zoom out the entities in screen. Scrolling the middle key in mouse can also zoom in or zoom out the entities. Use space key to pause/resume the simulation. Use direction keys to move the entities vertically or horizontally in screen. The simulation data can also be visualized in steps for inspection of the agent features in details, and users could press <Home/End> to forward or rewind the simulation process in steps, an*

*alternative method is clicking the left or right key in the mouse to move forward or rewind the process in steps. The visualization of the output binary data is encapsulated in a function as called visualizeAgent() in draw_func.py, and this function is also reused in our other programs such as evac-prt5-tool. Besides, when running the simulation with default settings, users also obtain another special form of data, which is suffixed by prt5. This special form of binary data is compatiable to FDS binary data for particles, and it can be used for 3D visualization by smokeview (NIST's 3D visualizer for FDS. In order to generate a smokeview script for 3D visualization, users should use another program called evac-prt5-tool to generate a script for smokeview.*

*In addition users could also check certain forces for each agent, mainly including the driving force* $\boldsymbol{f}_i^{drv}$ *and social force* $\sum_j \boldsymbol{f}_{ij}$. *These two types of forces are of theoretical interest, and thus we especially list them in the left bottom part of the pygame screen as shown in Figure 19. Here the driving force* $\boldsymbol{f}_i^{drv}$ *for agent 2 is 45.892N as indicated by motiveF, and the social force* $\sum_j \boldsymbol{f}_{ij}$ *is 87.571N as indicated by socialF.*

*Three lists as introduced before are also illustrated in Figure 19, which are see list, attention list and talk list. In Figure 19 the see list of agent 2 includes all of other agents except agent 0 and agent 1. In particular, agent 2 especially casts his or her attention to agent 3, and further talks with agent 3.*

*In the latest version of CrowdEgress program, the binary data can be visualized by Smokeview (3D visualizer for FDS) as shown in Figure 20. Smokeview is developed based on OpenGL, and it is also official 3D visualizer for FDS. In order to generate the script for Smokeview for CrowdEgress, users should combine the compartment layout (from a FDS Input files) with CrowdEgress output binary data. Another program is thus being developed, and it is called evac-prt5-tool, which is functional to combine FDS with CrowdEgress output data. This program of evac-prt5-tool is being tested, and we will integrate it into our CrowdEgress program in the near future. The latest version of evac-prt5-tool is uploaded in our website on sourceforge and github.*

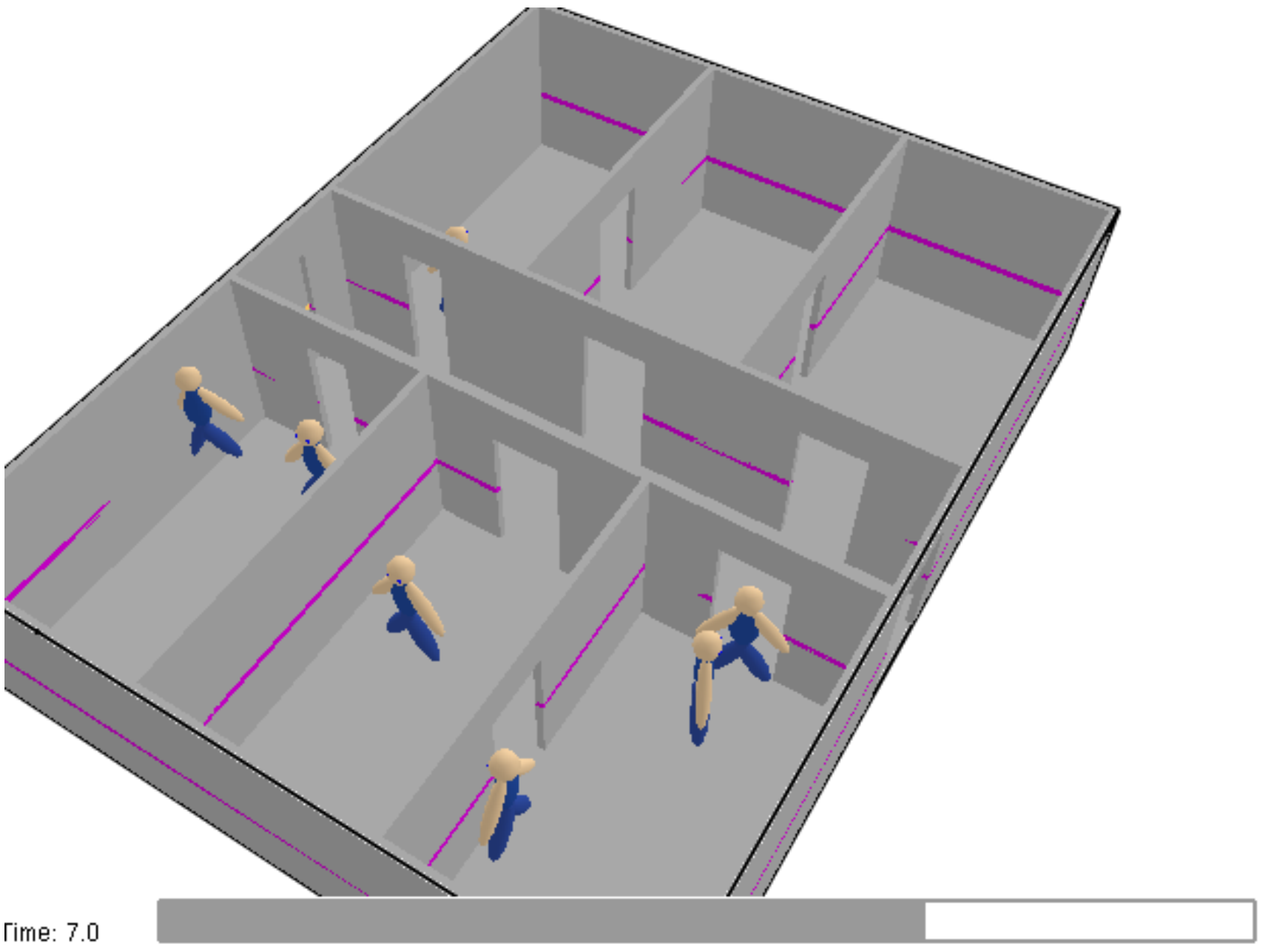

***Figure 20. Visualization of Simulation Output Data by Smokeview***

*There is also an array called PArray as listed in Figure 19. This array indicates how much an individual agent's opinion will be impacted by other agents in his or her attention list. In Table 7, for example, the Parray shows that agent 2's opinion is impacted by agent 3. In a probability sense agent 2 will keep balance between his or her own opinion and others' opinions with probability of 0.7 and 0.3. PArray will be timely updated as agents move and interact. How PArray is computed is mainly determined by the initial value of p and pMode parameter as specified in each agent's feature. Different agents may have different methods to update the p value, such as the random method or choosing a fixed value of p. The decision balance parameter p also critically affects agents' exit-selection behavior, and this feature is mainly studies in the panel of <DataTool> as shown below.*

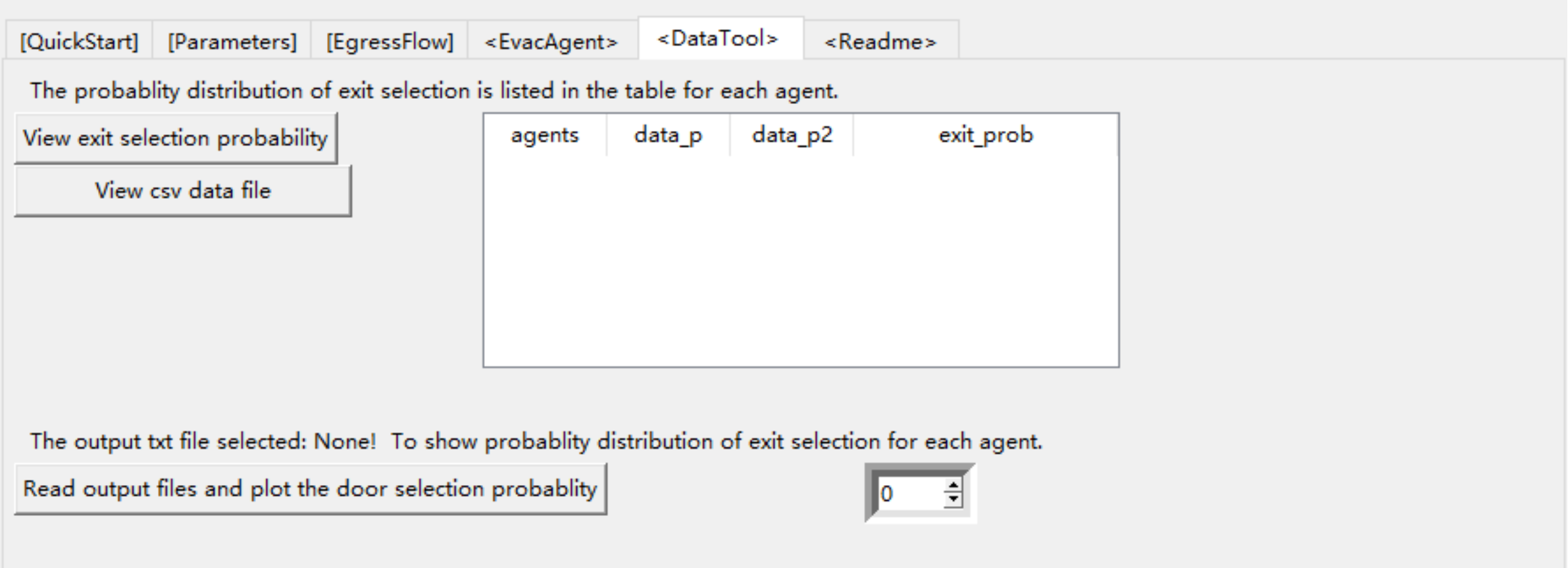

***Figure 21. User Interface of CrowdEgress (DataTool)***

*<View exit selection probability>: This button will list the initial value of decision balance parameter p as well as the exit-selection probability in the table on the right side.*

*<View csv data file>: This button will launch a treeview-based csv editor integrated in the GUI, and users can use it to modifly and save the input csv file as selected in the panel of QuickStart. As mentioned in Section 2, a major advantage of using this csv editor is that no empty cells will be added at the end of each line. As below we demonstrate this simple csv editor in Figure 22.*

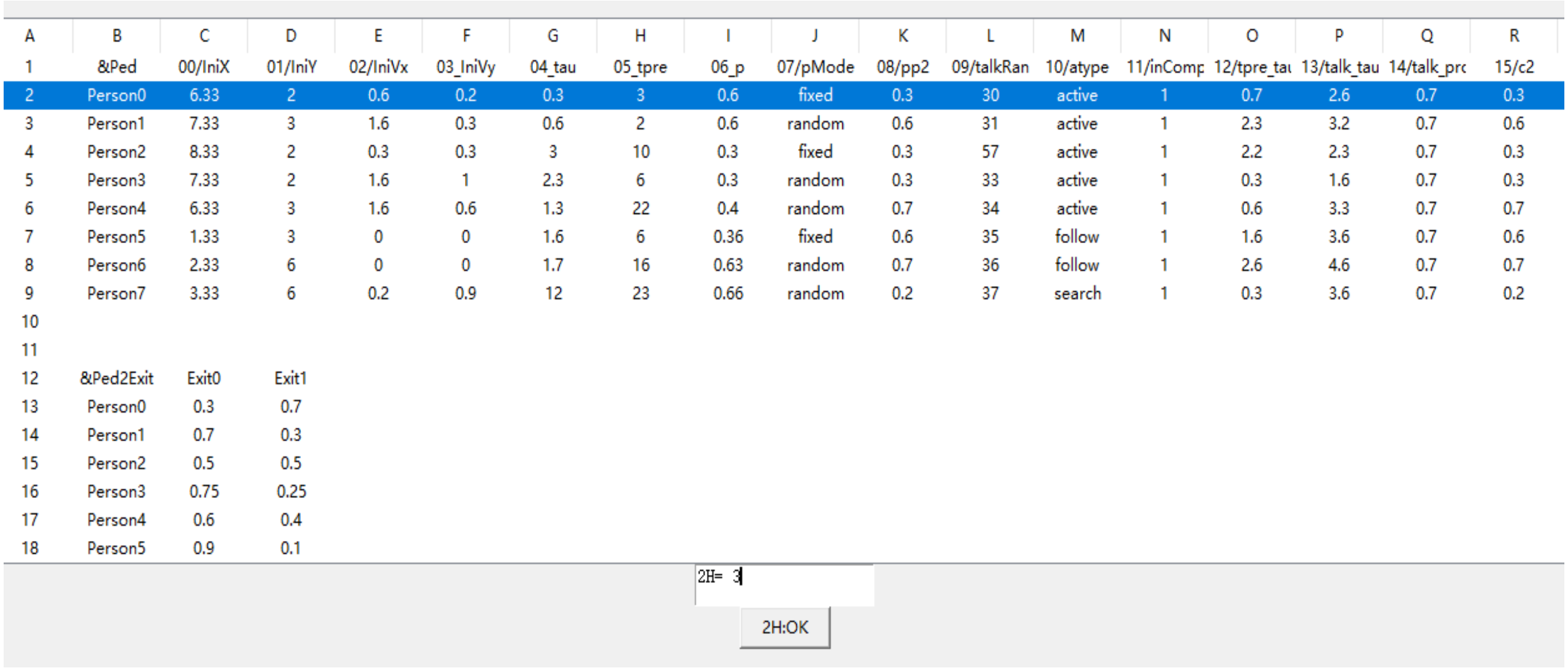

| A | B | C | D | E | F | G | H | I | J | K | L | M | N | O | P | Q | R |
|---|---|---|---|---|---|---|---|---|---|---|---|---|---|---|---|---|---|
| 1 | &Ped | 00/IniX | 01/IniY | 02/IniVx | 03_IniVy | 04_tau | 05_tpre | 06_p | 07/pMode | 08/pp2 | 09/talkRan | 10/atype | 11/inComp | 12/tpre_tau | 13/talk_tau | 14/talk_prc | 15/c2 |
| 2 | Person0 | 6.33 | 2 | 0.6 | 0.2 | 0.3 | 3 | 0.6 | fixed | 0.3 | 30 | active | 1 | 0.7 | 2.6 | 0.7 | 0.3 |
| 3 | Person1 | 7.33 | 3 | 1.6 | 0.3 | 0.6 | 2 | 0.6 | random | 0.6 | 31 | active | 1 | 2.3 | 3.2 | 0.7 | 0.6 |
| 4 | Person2 | 8.33 | 2 | 0.3 | 0.3 | 3 | 10 | 0.3 | fixed | 0.3 | 57 | active | 1 | 2.2 | 2.3 | 0.7 | 0.3 |
| 5 | Person3 | 7.33 | 2 | 1.6 | 1 | 2.3 | 6 | 0.3 | random | 0.3 | 33 | active | 1 | 0.3 | 1.6 | 0.7 | 0.3 |
| 6 | Person4 | 6.33 | 3 | 1.6 | 0.6 | 1.3 | 22 | 0.4 | random | 0.7 | 34 | active | 1 | 0.6 | 3.3 | 0.7 | 0.7 |
| 7 | Person5 | 1.33 | 3 | 0 | 0 | 1.6 | 6 | 0.36 | fixed | 0.6 | 35 | follow | 1 | 1.6 | 3.6 | 0.7 | 0.6 |
| 8 | Person6 | 2.33 | 6 | 0 | 0 | 1.7 | 16 | 0.63 | random | 0.7 | 36 | follow | 1 | 2.6 | 4.6 | 0.7 | 0.7 |
| 9 | Person7 | 3.33 | 6 | 0.2 | 0.9 | 12 | 23 | 0.66 | random | 0.2 | 37 | search | 1 | 0.3 | 3.6 | 0.7 | 0.2 |
| 10 | | | | | | | | | | | | | | | | | |
| 11 | | | | | | | | | | | | | | | | | |
| 12 | &Ped2Exit | Exit0 | Exit1 | | | | | | | | | | | | | | |
| 13 | Person0 | 0.3 | 0.7 | | | | | | | | | | | | | | |
| 14 | Person1 | 0.7 | 0.3 | | | | | | | | | | | | | | |
| 15 | Person2 | 0.5 | 0.5 | | | | | | | | | | | | | | |
| 16 | Person3 | 0.75 | 0.25 | | | | | | | | | | | | | | |
| 17 | Person4 | 0.6 | 0.4 | | | | | | | | | | | | | | |
| 18 | Person5 | 0.9 | 0.1 | | | | | | | | | | | | | | |

***Figure 21. CSV Data Editor Integrated in GUI (DataTool)***

*<Read output files and plot the door selection probability>: This button will allow users to select an output text file after the simulation is complete, and the program will read data from the text file and plot how the exit selection probability is dynamically updated through the simulation. The spinbox on the left side enables users to select an exit, of which the probability will be plotted, and the spinbox simply shows the index number of the exit. This index number starts from 0, and users could find this index number in testGeom screen.*

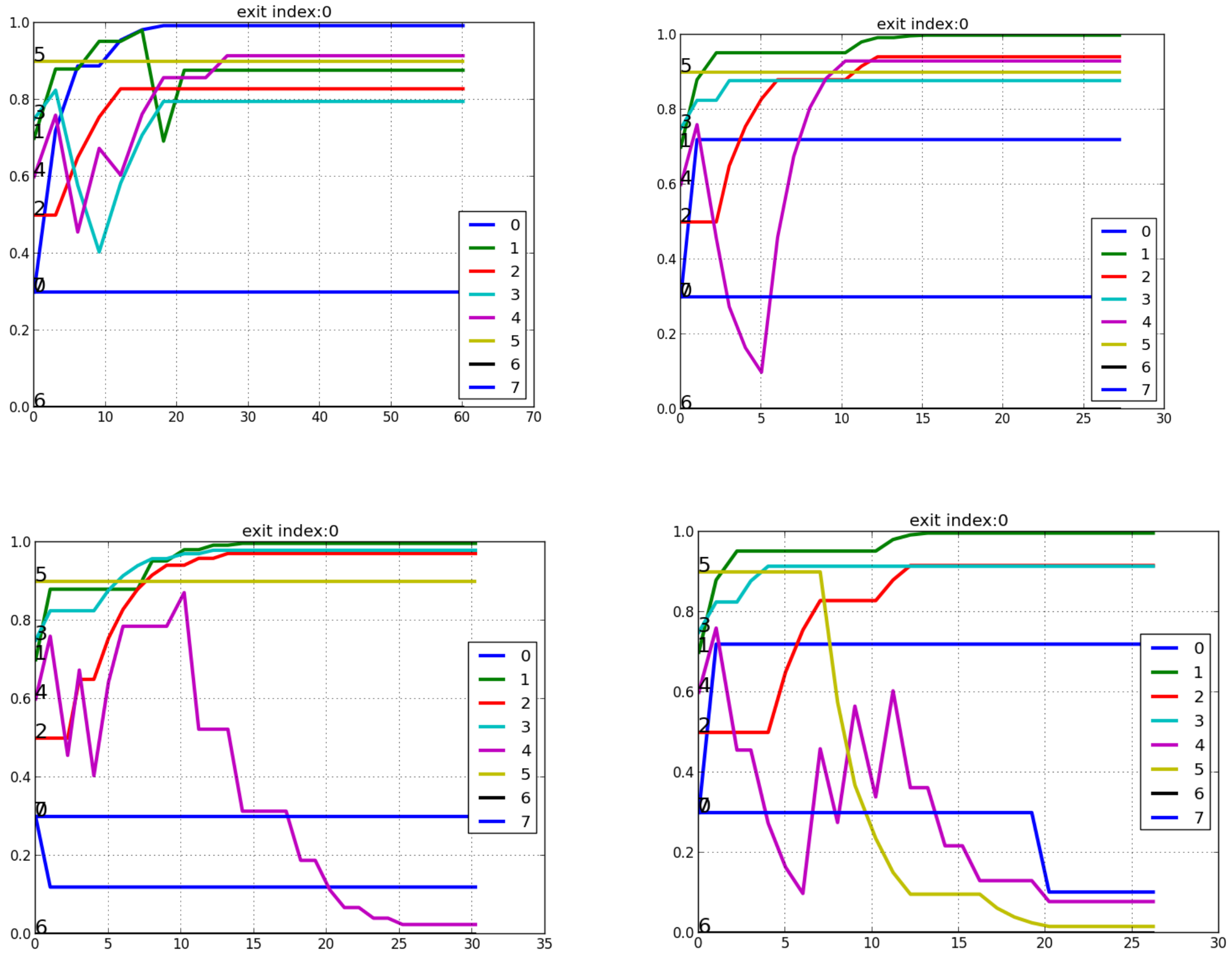

***Figure 23. Visualization of Exit Selection Probability***

*After the output text file is selected, the data will be extracted to show how the exit selection probability evolves for each individual agent through the simulation process. It is shown in Figure 23 that exit selection probability could either converge or be polarized as two groups. For example, the probability for agent 4 to use exit0 is initialized with 0.6, and throughout the simulation process the probability changes as the agent preference shifts from exit0 to exit1. Thus, the figure demonstrates how people are inclined to choose different exits in path-finding activities.*

## *6. Writing the Configuration File (config.txt) and Trying Examples*

*Configuration parameters of the simulation object could be alternatively read from a plain-text configuration file, as called config.txt, and it is to be placed in the same folder as the input csv files. In the following we will briefly introduce how to prepare config.txt. Especially, if users do not want to use GUI, for example in a small Linux system (e.g., Raspberry Pi OS), they can use main.py to launch the program without using any graphical user interface. The script of main.py follows a structured sequential workflow:*

- ***Initialization & Logging:*** *It starts by configuring a basic logging system to write debug, info, and warning messages to a file named log_examp.log.*
- ***Command-Line Interface (CLI) Parsing:*** *It uses sys.argv to accept input parameters from the command line, either for Microsoft Windows, MacOS or any Linux Distributions. Implementation principle is*
  - ***1 Parameter (****len(argv)==2****):*** *Expects a single csv input file (The agents and compartment layout should be specified in the csv file).*
  - ***2 Parameters (****len(argv)==3****):*** *Expects a .csv file and an fds file (The agents should be specified in the csv file while a single-floor compartment layout should be included in the fds file).*
  - ***3+ Parameters (****len(argv)>3****):*** *Rejects the input as invalid and terminates the program.*

*In order to properly use main.py to launch the program, users should first write a config.txt to configure the simulation object properly. Otherwise, the simulation will be configured by using the default values in __init__() in simulation.py. The method readconfig() in simulation.py is designed to parse config.txt and load its key-value pairs to configure the simulation object.*

*<u># The parameters to configure simulation window (pygame screen).</u>*
*# User may also use pageup/pagedown to zoom in/out the screen, and use the direction keys to move entities in the screen.*
*ZOOM=20.0*
*xSpace=30.0*
*ySpace=60.0*

*# The parameter to configure the egress flow solver.*
*# xpt and ypt are the number of points in x and y axises.*
*# xmin and xmax specify the range of flow field in x axis; # ymin and ymax specify the range of flow field in y axis*
*xmin=0*
*xmax=10*
*ymin=0*
*ymax=10*
*xpt=60*
*ypt=30*

*# Solver: 0 No egress flow field; 1 Nearest exit flow field; 2 Exit selection flow field*
*solver=1*

*# Parameters to configure the simulation object.*
*# Groupbehavior: 0 No social group effect; 1 Compute social group effect*
*# dumpBinary: 0 No output data; 1 Dump and save simulation data in binary file*

*groupbehavior = 1*
*pre-evac = 1*
*dumpBinary = 1*

***<u>Try Examples:</u>***

*There are currently several examples in the repo of <u>https://sourceforge.net/p/crowdegress</u> or <u>https://github.com/godisreal/crowdEgress</u>. For instance there are standard example of a single room with two exits or three exits, which are basically used to test how agents socially interact and select different exits. The example of a room with one exit is also widely used to test crowd behavior at bottlenecks such as the faster-is-slower effect (Helbing, Farkas, Vicsek, 2000).*

*There are also some more complicated examples. A typical one is learned and extended from MassEgress project (Pan, 2006), and another one is from our IEEE Conference paper (Wang et. al., 2008). Some FDS+Evac test cases are also included. Users can also learn how to write the csv files from the examples.*

*In order to run such complicated examples it is suggested that users should first check the mesh parameters. Especially, the total number of x point and y points are to be properly given to build a mesh for flow computation. The flow solver may take some computational time to generate the egress flow field, especially if x point and y point are relatively large to generate a refined mesh.*

## *Acknowledgments*

*The authors are thankful to Bo Xiong and Vivek Kant for helpful comments on earlier work in University of Connecticut. The author appreciates the research program funded by NSF Grant # CMMI-1000495 (NSF Program Name: Building Emergency Evacuation - Innovative Modeling and Optimization).*

*The authors also thank GeögerTopi for sharing his wonderful python script on Discussion Forum of FDS, and thank Salah Benkorichi for sending the information to me! I mainly modified xyz.shape = (7,nplim) for evac binary data. The authors also thank Shen Shen for sharing his repo of social force model on Github. His wonderful codework inspires us to start this python-based program of crowdEgress at early stage.*

*The authors also thank Gregor Jäger and Laura Pituková for assistance in reviewing the codework.*

*The program source code and numerical test cases are mainly included online at https://sourceforge.net/p/crowdegress/discussion/general/. If you have any comment or inquiry about the testing result, please feel free to contact me at wp2204@gmail.com or start an issue on the repository.*